\documentclass[aps,epsfig,showpacs]{revtex4}
\usepackage{graphicx}
              
\def\be{\begin{equation}}
\def\ee{\end{equation}}
\def\bfi{\begin{figure}}
\def\efi{\end{figure}}
\def\bea{\begin{eqnarray}}
\def\eea{\end{eqnarray}}

\begin{document}

\title{Fluctuations of two-time quantities and non-linear response functions}

\author{F.Corberi}
\affiliation{Dipartimento di Matematica e Informatica and 
INFN, Gruppo Collegato di Salerno, and CNISM, Unit\'a di Salerno,
via Ponte don Melillo, Universit\`a di Salerno, 84084 Fisciano (SA), Italy}
\author{E. Lippiello}
\affiliation{Dipartimento di Scienze Ambientali, Seconda Universit\'a di Napoli, 
Via Vivaldi, Caserta, Italy}
\author{A.Sarracino}
\affiliation{Dipartimento di Matematica e Informatica,
 via Ponte don Melillo, Universit\`a di Salerno, 84084 Fisciano (SA), Italy}
\author{M. Zannetti}
\affiliation{Dipartimento di Matematica e Informatica,
 via Ponte don Melillo, Universit\`a di Salerno, 84084 Fisciano (SA), Italy}

 \date{\today}

\begin{abstract}

We study the fluctuations of the autocorrelation and autoresponse functions and,
in particular, their variances and co-variance.
In a first general part of the Article, 
we show the equivalence of the variance of the response function 
with the second-order susceptibility of a composite operator, 
and we derive an equilibrium fluctuation-dissipation theorem beyond-linear order
relating it to the other variances.
In a second part of the paper we apply the formalism to the study to
non-disordered ferromagnets, in equilibrium or in the coarsening kinetics
following a critical or sub-critical quench. We show numerically that
the variances and the non-linear susceptibility obey 
scaling with respect to the coherence length $\xi$ in equilibrium,
and with respect to the growing length $L(t)$ after a quench, similarly
to what is known for the autocorrelation and the autoresponse functions. 
   
\end{abstract}


\maketitle


\section{Introduction}\label{intro}

Two-time quantities, such as the autocorrelation
function $C(t,t_w)$ and the associated linear 
response function $\chi (t,t_w)$,
describing the effects of a perturbation, are generally considered in
experiments, theories and numerical investigations.
In equilibrium the fluctuation-dissipation theorem (FDT) holds,
providing an important tool to
study coherence lengths and relaxation times by means of susceptibility
measurements.

Beside equilibrium, 
the pair $C$ and $\chi $ has been thoroughly investigated 
also in slowly relaxing systems,
among which supercooled liquids, glasses, spin-glasses and quenched
ferromagnets, as natural quantities to characterize and
study the aging behavior. 
In this context, the fluctuation-dissipation ratio $X(t,t_w)=d\chi/dC$
was defined~\cite{x} in order to
quantify the {\it distance} from equilibrium, where $X\equiv 1$.
Particularly relevant is its limiting value
$X_\infty= \lim _{t_w\to \infty} \lim _{t\to \infty} X(t,t_w)$ 
due to its robust universal properties~\cite{numxinfty,godluck,analxinfty,noiteff}.
Complementary to $X$, the concept of an effective temperature $T_{eff}=T/X$,
has been thoroughly applied in several contexts~\cite{cukupe}, although its
physical meaning has not yet been completely clarified.
Moreover, the fluctuation-dissipation ratio was also proved~\cite{fmpp} to be related to  
the overlap probability distribution of the equilibrium
state at the final temperature of the quench, providing an important bridge between equilibrium
and non-equilibrium. 
Finally, in the context of coarsening systems, 
the behavior of the response function was shown to be
strictly linked to geometric properties of the interfaces~\cite{rough,rough1},
allowing the characterization of their roughness, and, in the
case of phase-ordering on inhomogeneous substrates, 
to important topological properties of the underlying 
graph~\cite{graphs}.
  
Besides this manifold interest in average two-time quantities,
more recently considerable
attention has been paid also to the study of 
their local fluctuations,
which are now accessible in large-scale
numerical simulations~\cite{numfluc} and, due to new techniques, also 
in experiments~\cite{2-10diarxiv0401326}.    
The reasons for considering these quantities are various:

- In disordered systems, since averaging over the disorder makes the usual two-particle
correlation function (structure factor) short ranged even in those cases
where a large coherence length $\xi $ is present, quantities related
either to the spatial fluctuations of $C$~\cite{c4,BB,c4chi22} or  
to non-linear susceptibilities~\cite{BB,non_lin1&2} have been proposed 
to detect and quantify $\xi $. 

- Local fluctuations of two-time quantities 
are associated with the dynamical heterogeneities observed in several
systems which are believed to be a key for local rearrangements
taking place in slowly evolving systems~\cite{11-21diarxiv0401326,numfluc}. 
In the context of spin models,
it was shown~\cite{numtris} 
that these fluctuations can be conveniently used to highlight 
the heterogeneous nature of the system.

- In~\cite{arxiv/0109150} 
it was shown that in a large class of 
glassy models the action  describing the asymptotic dynamics is invariant under the 
transformation of time $t\to h(t)$, denoted as time re-parametrization. 
This symmetry is expected to hold true in
glassy systems with a finite effective temperature but not
in coarsening systems, where $T_{eff}=\infty $~\cite{nontris}.
Then, restricting to glassy systems, it was proposed~\cite{arxiv/0109150,numtris} 
that the aging kinetics could be 
physically interpreted as the coexistence of different
parametrization $t\to h_r(t)$ slowly varying in space ${\bf r}$.
According to this interpretation, spatial fluctuations of two-time quantities 
should span the possible
values of $C$ and $\chi$ associated to different choices of $h(t)$.
Since the correlation and the response function transform 
in the same way under the time-re-parametrization transformation,
the same curve 
$\chi (C)$ relating the average quantities is expected to hold also for 
the fluctuations.  
This property was proposed in~\cite{arxiv/0109150,numtris} 
as a check on the time-re-parametrization
invariance, and the results tend to conform to this interpretation.

- In~\cite{corbandolo} it was claimed that, at least in the 
context of non-disordered coarsening systems, 
fluctuations of two-time quantities
encode the
limiting fluctuation-dissipation ratio $X_\infty $, similarly to the
fluctuation-dissipation relation between the fully averaged 
quantities $\chi $ and $ C $.

In the first part of this paper, we discuss the  definition of the fluctuating versions 
$\widehat C_i, \widehat \chi _i$ of $C_i$ and $\chi _i$ on site $i$, and 
consider their (co-)variances $V^C_{ij}=\langle \widehat C_i \widehat C_j\rangle-C_iC_j$, $V^\chi _{ij}$,
and $V^{C\chi}_{ij}$ (defined analogously to $V^C_{ij}$).
We present a rather detailed and complete study 
of these quantities and their relation
with a non-linear susceptibility ${\cal V}^\chi_{ij}$ (defined in Eq. \ref{chi2r2})
related to the
fluctuations of $\widehat \chi _i$ introduced in~\cite{non_lin1&2}.
We show that, for $i\neq j$,
the variance $V_{ij}^{\chi}$ of $\widehat \chi _i$ is equal to ${\cal V}_{ij}^\chi$. 
This allows us to derive a relation between  
$V_{ij}^C$, ${\cal V}_{ij}^\chi$ and $V_{ij}^{C\chi}$, 
which can be regarded as a second order fluctuation-dissipation theorem
(SOFDT) relating these quantities in equilibrium. 
The SOFDT holds for every choice of $t$ and $t_w$ and 
of $i,j$ and is completely general for Markov systems.
It represents also a relation between
the second moments of $\widehat C_i$ and $\widehat \chi_j$ for $i\neq j$,
but not for $i=j$ because, 
in this case, ${\cal V}_{ij}^\chi$ 
cannot be straightforwardly interpreted as a variance.
Prompted by the SOFDT,
we argue that ${\cal V}_{ij}^\chi$, rather than $V_{ij}^{\chi}$, is the 
natural quantity to be considered, on an equal footing 
with the variances $V_{ij}^C$ and $V_{ij}^{C\chi}$, 
to study scaling behaviors, and 
to detect and quantify correlation lengths. Being a susceptibility,
${\cal V}$ could in principle be accessible in experiments.

These ideas are tested in
the second part of the paper, where we study numerically the behavior
of $V_{ij}^C$, $V_{ij}^{C\chi}$ and of ${\cal V}_{ij}^\chi$
in non-disordered
ferromagnets in equilibrium or in the non-equilibrium kinetics
following a quench to a final temperature $T$ at or below $T_c$. 
Restricting to the cases with $T\geq T_c$ the same problem has been
recently addressed analytically by Annibale and Sollich~\cite{annibale} 
in the context of the soluble spherical model. Here we 
carry out the analysis 
in the finite-dimensional Ising model, focusing particularly
on the scaling properties.
Focusing on the $k=0$ Fourier component 
$V^C_{k=0}=(1/N)\sum _{i,j=1}^N V^C_{ij}$ (and similarly for
the other quantities)
our results show a pattern of behaviors
for $V^C_{k=0}$, $V^{C\chi}_{k=0}$ and ${\cal V}^\chi_{k=0}$ 
similar to what is known for
$C$ and $\chi $.
In particular, in a quench at $T_c$, one finds the asymptotic scaling form
$V^C_{k=0}(t,t_w)\propto V^{C\chi}_{k=0}(t,t_w)\propto {\cal V}^\chi _{k=0}(t,t_w)\propto t_w^{b_c}f(t/t_w)$,  
where the exponent 
$b_c=(4-d-2\eta )/z_c$ can be expressed in terms 
of the equilibrium static and dynamic critical exponents $\eta$ and $z_c$,
in agreement with what was found in~\cite{annibale}.
In quenches below $T_c$, in the time sector with $t_w\to \infty$ and
$t/t_w=const.$, usually referred to as aging regime, 
we find a scaling form
$V^C_{k=0}(t,t_w)=t_w^{a_C}f(t/t_w)$ (and similarly for $V^{C\chi}_{k=0}$ 
and ${\cal V}^\chi _{k=0}$),
where, in contrast to the critical quench, 
$a_C$ and $f$ are genuinely non-equilibrium
quantities that cannot be straightforwardly related to
equilibrium behaviors.  
 
Our results allow us to discuss also the issue of a direct correlation between
the fluctuating parts of $C$ and $\chi$, as predicted for glassy systems
by the time-re-parametrization invariance scenario.
In the aging dynamics of coarsening systems, 
we find that for large $t/t_w$ the ratio
${\cal V}^\chi/V^C$ diverges, both in the quench at $T_c$
and below $T_c$. This implies that
$\widehat C$ and $\widehat \chi$ 
are not related as to follow the curve 
$\chi (C)$, in contrast with the above mentioned scenario.  

This paper is organized as follows: in Sec.~\ref{defin} we 
introduce and discuss general definitions of 
the fluctuating quantities,
their variances and co-variances; in Sec.~\ref{eqrel} we discuss 
the relations among them and with the second-order susceptibility ${\cal V}^\chi$. 
In Sec.~\ref{ferro} we specialize the above concepts to the
case of ferromagnetic systems. We study
the behavior of $V^C$, $V^{C\chi}$ and ${\cal V}^\chi$ in equilibrium in Sec.~\ref{equiferro},
relating their large $t-t_w$ behavior to the coherence length in Sec.~\ref{static}.
The non-equilibrium kinetics is considered in Sec.~\ref{quench}:
critical quenches are studied in 
Sec.~\ref{critque}, while Sec.~\ref{beque} is devoted to sub-critical quenches.  
The results of these Sections are related to the issue of time-re-parametrization
invariance in Sec.~\ref{time}.
Last, in Sec.~\ref{concl} we summarize, draw our conclusions and discuss some
open problems and perspectives. Four appendixes contain some technical points. 
 
\section{Fluctuating quantities and variances} \label{defin}

Let us consider a system described by a set of variables $\sigma _i$
defined on lattice sites $i$. 
In order to fix the notation we consider discrete variables, referred
to as spins,
the evolution of which is described by a master equation. The results of this
paper, nonetheless, apply as well to continuous variables subjected
to a Langevin equation (specific differences between the two cases
will be noticed whenever the case).
The auto-correlation function is defined as
\begin{equation}
C_{i}(t,t_w)=\langle\sigma_i(t)\sigma_i(t_w)\rangle
-\langle\sigma_i(t)\rangle \langle\sigma_i(t_w)\rangle .
\label{1.1}
\end{equation}
Using the symbol $\hspace{1mm} \widehat{} \hspace{1mm}$ to denote 
the fluctuating quantities whose average
gives the usual functions, one has
$\widehat C_i(t,t_w)=[\sigma _i(t)-\langle\sigma _i(t)\rangle]
[\sigma_i(t_w)-\langle\sigma_i(t_w)\rangle]$.
The (auto-)response function is defined as
\begin{equation}
\chi_{i}(t,t_w)=T\int_{t_w}^t dt'\left. \frac{\delta \langle \sigma _i(t)\rangle _h }{\delta h_i(t')}\right |_{h=0},
\label{1.2}
\end{equation}
where $\langle \ldots \rangle_h$ means an average over a process where an impulsive perturbing
field $h$ has been switched on at time $t'$.
Notice that
a factor $T$ has been included in the definition~(\ref{1.2}) of the response.
The presence of the derivative in Eq. (\ref{1.2}) makes a definition of a fluctuating
part of $\chi _i$ not straightforward in the case of discrete variables 
(see Appendix I for a discussion of a possible definition of 
$\widehat \chi _i$ based on the definition (\ref{1.2}), where the perturbation $h_i$ is
present).
This problem can be bypassed using an out-of-equilibrium fluctuation-dissipation
relation
\be
\chi _i(t,t_w)=\langle \widehat \chi _i(t,t_w)\rangle,
\label{fdt2}
\ee
where in the limit of
vanishing $h$ the derivative of Eq.~(\ref{1.2}) is worked out analytically, and
on the right hand side appear specific correlation functions (see e.g. Eq.~(\ref{1.5}) and
discussion below)
computed in the unperturbed dynamics. Such a relation has been obtained in different 
forms in~\cite{rlinsoft,Crisanti,chatelain,rlin,Diez,Berthier,non_lin1&2,Baiesi}.
This allows one to introduce a fluctuating part of the susceptibility defined over
an unperturbed process.
Eq. (\ref{fdt2}) is at the basis of the so called {\it field-free} methods for the 
computation of response functions allowing the computation of $\chi _i$ 
without applying any perturbation.

With the quantities introduced above, one
can build the following (co-)variances 
\begin{eqnarray}
V^C_{ij}(t,t_w)&=&\langle\widehat {\delta C_{i}}(t,t_w)\widehat {\delta C}_{j}(t,t_w)\rangle \label{1.6}\\
V^{\chi} _{ij}(t,t_w)&=&\langle\widehat {\delta \chi}_i(t,t_w)\widehat {\delta \chi} _j(t,t_w)\rangle
\label{1.71}\\
V^{C\chi} _{ij}(t,t_w)&=&\langle \widehat {\delta C_{i}}(t,t_w)\widehat {\delta \chi} _j(t,t_w)\rangle
\label{1.72}
\end{eqnarray}
where, for a generic observable $A$, we have defined $\widehat {\delta A} \equiv \widehat{A}-\langle \widehat A \rangle$. 
Notice that we restrict the analysis to variances obtained by taking products of two-time quantities
on different sites but with the same choice of times $t,t_w$. 
$V^C_{ij}(t,t_w)$ is the 4-point
correlation function introduced in~\cite{c4} to study cooperative effects
in disordered systems, usually denoted as $C_4$.

As discussed in \cite{rlin,algoritmi}, for a given unperturbed model, there are
many possible choices of the perturbed transition rates, which give rise to
different expressions for $\widehat \chi _i$. However, as shown in \cite{algoritmi},
and further in Appendix I, we expect all these choices to lead to 
approximately the same values of the variances
introduced above (with the notable exception of the equal site variance $V^{\chi}_{ii}$, which,
however, is not of interest in this paper). Then, in the following, we will consider the 
expression
\begin{equation}
\widehat \chi_i(t,t_w)=
\frac{1}{2}\left[\sigma _i(t)\sigma_i(t)- \sigma _i(t)\sigma_i(t_w)- 
\sigma_i(t)  \int _{t_w}^t dt_1 B_i(t_1)\right ],
\label{1.5}
\end{equation}
where $B_i=-\sum _{\sigma '}[\sigma _i-\sigma' _i] 
w(\sigma '\vert \sigma)$, $w(\sigma '\vert \sigma)$ 
being the transition rate for going from the configuration $\sigma $ to $\sigma '$.
This form has been obtained in~\cite{rlin} (and,
in an equivalent formulation,
for continuous variables in~\cite{rlinsoft,non_lin1&2}).

The relation~(\ref{fdt2}) with the choice (\ref{1.5}) has the advantage of a large generality, holding for Markov processes 
with generic unperturbed transition rates, both for continuous and discrete variables.
Other possible relations between the response and quantities computed on unperturbed trajectories
have been proposed~\cite{Crisanti,chatelain,Diez,Berthier} but we do not consider them here because,
as discussed in~\cite{algoritmi},
in those approaches either the response is not related to
correlation functions of observable quantities in the unperturbed system, as in~\cite{Crisanti,Diez,Berthier},
or, in the case of Ref.~\cite{chatelain}, it is restricted to a specific
systems (Ising) with a specific (Heat bath) transition rate.

The $k=0$ Fourier component of the correlation and 
response functions are usually considered 
to extract physical information, such as spatial coherence or
relaxation times, from the (unperturbed) system under study.
The $k=0$ mode $V^C_{k=0}(t,t)$ of the variance of $\widehat C$, defined through
\be
V^C_{k=0}(t,t_w)=\frac{1}{N}\sum _{i,j=1}^N V^C _{ij}(t,t_w),
\label{keq0}
\ee
has been considered to access the same information in disordered systems.
This might suggest that the same information is contained
in the $k=0$ component of the other variances. Notice that, for $V^{\chi}$,
the sum~(\ref{keq0}) includes the equal site term $V^{\chi}_{ii}$ which,
as anticipated, takes different values according to the
specific choices of the fluctuating part of the response.
We will deal with this problem later.

\section{Equilibrium relation between variances 
and non-linear susceptibilities} \label{eqrel}

In this section we derive a relation between the variances and the
non-linear susceptibility ${\cal V}^\chi$ (defined in Eq. \ref{chi2r2}) that will be 
interpreted as a second-order
fluctuation-dissipation theorem (SOFTD) relating these quantities. 
We sketch here the basic results,
further details and formalism are contained in Appendix II.

Let us start by recovering the usual FDT.
In equilibrium, using time translation and time inversion invariance, 
namely the Onsager relations,
it can be shown~\cite{non_lin1&2} that 
\begin{equation}
\langle\sigma_i(t) B_i(t_1)\rangle_{eq}=
-\frac{\partial}{\partial t_1}\langle\sigma_i(t)\sigma_i (t_1)\rangle_{eq},
\label{2.1}
\end{equation}
valid for $t>t_1$. Plugging this relation into Eqs.~(\ref{fdt2},\ref{1.5}) one retrieves the
usual fluctuation-dissipation theorem
\begin{equation}
\langle \widehat D_i(t,t_w)\rangle=0,
\label{2.5}
\end{equation}
where we have introduced the quantity
\be
\widehat D_i(t,t_w)=\widehat \chi_i (t,t_w)+\widehat C_i(t,t_w)-\widehat C_i(t,t).
\label{hatd}
\ee
Notice that, for Ising spins $\sigma _i=\pm 1$, $\widehat C_i(t,t)\equiv 1$ and does not fluctuate.

The next step is to seek for a relation holding between the variances.
Since the mechanism whereby this relation is obtained is different
for equal or different sites $i,j$ (due to the sensitivity of $V^{\chi}_{ii}$ to the
choice of $\widehat \chi _i$), we split the arguments into separate
sections.

\subsection{$\mathbf {i\neq j}$} \label{ineqj}

Defining the second moment of $\widehat D_i$ as 
$V^D_{ij}(t,t_w)=\langle \widehat {\delta D}_i(t,t_w)\widehat {\delta D}_j(t,t_w)\rangle$, 
and using the equilibrium property (\ref{2.1}) it is easy to show that
\be
V^D_{ij}(t,t_w)=
V^{\chi} _{ij}(t,t_w)+2V^{C\chi}_{ij}(t,t_w)+V^C_{ij}(t,t_w)-V^C_{ij}(t,t).
\label{vd}
\ee
Proceeding in a similar way as done in the derivation of Eq.~(\ref{2.5}),
in Appendix II we show that, for $i\neq j$, the r.h.s. of Eq. (\ref{vd}) vanishes 
in equilibrium. Hence we have the following SOFDT
\be
V^D_{ij}(t,t_w)=0.
\label{1quadratoa}
\ee
This relation holds for every choice of the fluctuating part
of $\chi $: Indeed, we have already noticed that on different sites $i,j$
the variances involved in the r.h.s. of Eq.~(\ref{vd}) are independent
on that choice.
Interestingly, Eq.~(\ref{1quadratoa}) shows that not only the first moment of
$\widehat D_i$ vanishes (due to the FDT~(\ref{2.5})), but also
the second moment.
Moreover, as shown in Appendix I,
the equal site variance
$V^D_{ii}$ is not zero (due to the divergence of the term $K_i^\chi$ (or $\tilde K_i^\chi$)
appearing in $V^\chi_{ii}$, see Eqs.~(\ref{kappa1}),~(\ref{divh})), indicating
that $\widehat D$ is not identically vanishing, and hence it is a truly fluctuating quantity.
This leads to the surprising conclusion that $\widehat D$ is 
an uncorrelated variable for any choice of $i,j$ and of $t,t_w$, and
in any equilibrium state of any Markovian model.
This observation, which might have far reaching consequences, will
be enforced in Sec.~\ref{ferro} to disentangle quasi-equilibrium 
correlation from the genuine non-equilibrium ones in aging systems.

\subsection{$\mathbf {i=j}$} \label{ieqj}

For $i=j$ a relation such as Eq.~(\ref{2.5}) cannot be satisfied for any
choice of the fluctuating part of $\chi $. 
In order to show that, let us first observe that,
recalling Eq.~(\ref{vd}), if Eq.~(\ref{1quadratoa}) were to hold
also for $i=j$, the quantity 
$-2V^{C\chi}_{ii}(t,t_w)-V^C_{ii}(t,t_w)+V^C_{ii}(t,t)$ should equal
$V^{\chi}_{ii}(t,t_w)$.
This quantity can be easily computed, yielding
\be
-2V^{C\chi}_{ii}(t,t_w)-V^C_{ii}(t,t_w)+V^C_{ii}(t,t)=
-\chi_i^2(t,t_w)-\Delta _i(t,t_w),
\label{hyp}
\ee
where $\Delta _i(t,t_w)=2\langle \widehat C_i(t,t_w)\widehat \chi_i(t,t_w)\rangle
+\langle \widehat C_i^2(t,t_w)\rangle-\langle \widehat C_i^2(t,t)\rangle$
is a quantity which vanishes for Ising spins, as can be easily shown using the 
definitions of $\widehat C_i$ and $\widehat \chi _i$ and the property~(\ref{propB1}).
On the other hand, computing $V^{\chi}_{ii}$ directly leads to
the result (see Appendix I)
\be
V^{\chi}_{ii}(t,t_w)=-\chi_i^2(t,t_w)-\Delta _i(t,t_w)+K_i^{\chi}(t,t_w), 
\label{bbbbb}
\ee
where $K_i$, given in Eq. (\ref{kappa1}),
is a quantity that has been studied in specific models in \cite{algoritmi} and found 
to be positive and diverging as $t-t_w$ increases.
Expression (\ref{bbbbb}) is different from the
r.h.s. of Eq.~(\ref{hyp}), thus proving that the SOFTD does not
hold for $i=j$. Worse, the quantity $K_i^{\chi}$
appearing in Eq.~(\ref{bbbbb}) prevents the possibility of any
direct relation between the variances because it introduces an 
explicit time-dependence.

\subsection{The non-linear susceptibility ${\mathbf{\cal V}^\chi _{ij}(t,t_w)}$}

In order to remove the asymmetry between $i=j$ and $i\neq j$ and proceed further,
the idea is to search for a quantity ${\cal V}^\chi_{ij}$ related to $V^\chi_{ij}$
such that ${\cal V}^\chi _{i\neq j}=V^\chi _{i\neq j}$, while on equal sites
the equilibrium value of ${\cal V}^\chi _{ii}$ equals the r.h.s. of Eq.~(\ref{hyp}).
This would allow one to arrive at
a pair of relations analogous to Eqs.~(\ref{1quadratoa},\ref{vd}) for any $ij$.
As shown in Appendix III,  
the second order susceptibility 
\be
{\cal V}^\chi_{ij}(t,t_w)\equiv \int_{t_w}^t dt_1\int_{t_w}^t dt_2 \left [R^{(2,2)}_{ij;ij}(t,t;t_1,t_2)
-R_i(t,t_1)R_j(t,t_2) \right ],
\label{chi2r2}
\ee
where 
\be
R^{(2,2)}_{ij;ij}(t,t;t_1,t_2)\equiv T^2\left .
\frac{\delta^2\langle\sigma_i(t)\sigma_j(t)\rangle_h}
{\delta h_{i}(t_1)\delta h_j(t_2)}\right|_{h=0}
\label{app1.2}
\ee
is the non-linear impulsive response function  
proposed in~\cite{non_lin1&2}
to study heterogeneities in disordered systems, meets the
requirements above.
Then, recalling Eq.~(\ref{hyp}), one has the relations
\be 
{\cal V}^D_{ij}(t,t_w)=0,
\label{quadratoa}
\ee
and
\be
{\cal V}^D_{ij}(t,t_w)=
{\cal V}^{\chi} _{ij}(t,t_w)+2V^{C\chi}_{ij}(t,t_w)+V^C_{ij}(t,t_w)-V^C_{ij}(t,t)
\label{quadratoa1}
\ee
formally identical to Eqs.~(\ref{1quadratoa},\ref{vd}), but holding
for every choice of the sites $i,j$ and hence also
for the $k=0$ component, namely
\be
{\cal V}^D_{k=0}(t,t_w)=0.
\label{quadrato}
\ee

In summary, one always has an equilibrium relation (Eq.~(\ref{quadratoa}) or~(\ref{quadrato}))
between the second order response defined in Eqs.~(\ref{chi2r2},\ref{app1.2})
and the variances $V^C_{ij}$ and $V^{C\chi}_{ij}$. In the case of different sites $i\ne j$,
this non-linear response is also the variance of 
$\widehat \chi $, whereas
on equal sites there is no analogous interpretation, and neither 
it is possible to obtain a relation involving directly $V^{\chi}_{ii}$. 

Coming back to the problem discussed at the end of
Sec. \ref{defin}, namely the possibility of extracting physical information
on the unperturbed system from the $k=0$ mode of the variances,
some considerations are in order.
First, it is clear that, concerning $V^\chi _{k=0}$, its value changes 
depending on the way the perturbation is introduced (via the term
$V^\chi _{ii}$). In this way this quantity mixes information regarding
the perturbation with those of interest. The quantity ${\cal V}^\chi _{k=0}$,
instead, does not suffer from this problem, since its equal site value can
always be related to quantities that do not depend on the choice of 
the perturbation. Moreover, for large times, $V^\chi_{k=0}$ 
(defined analogously to Eq.~(\ref{keq0})) turns out to be
dominated by the equal site contribution.
Indeed, whatever definition of $\widehat \chi _i$ is adopted,
either the quantity $K_i^\chi$ or $\tilde K_i^\chi$ come in 
(see Eqs.~(\ref{divh},\ref{kchi})),
which are either infinite ($\tilde K_i^\chi$)
or diverging with increasing $t-t_w$ ($K_i^\chi$).
These considerations suggest the use of ${\cal V}^\chi _{k=0}$.
Indeed it has been shown in specific cases~\cite{non_lin1&2} that
this quantity contains information on relevant properties, among which 
the coherence length, similarly to the variance $V^{C} _{ij}$ and
therefore has an important physical meaning.

\section {Fluctuations in Ferromagnets}\label{ferro}

Specializing the general definitions given above to the case
of ferromagnetic systems, in this Section we study
the behavior of the $k=0$ mode of the quantities introduced above in 
the Ising model in equilibrium (Sec.~\ref{equiferro})
and in the non-equilibrium kinetics following a quench to
$T_c$ (Sec.~\ref{critque}) or below $T_c$ (Sec.~\ref{beque}).  
Our main interest is in the scaling of these quantities
with respect to the characteristic length of the system.
From this perspective, it is quite natural to focus on
${\cal V}^\chi_{k=0}$ rather than on $V^{\chi}_{k=0}$. 
Indeed we will show that
in any case $V^C_{k=0}$, $V^{C\chi}_{k=0}$ and ${\cal V}^\chi_{k=0}$ obey scaling forms
from which a correlation length can be extracted.
On the other hand, as already anticipated, these scaling properties are masked in
$V^{\chi}_{k=0}$ 
by the term $K_i^\chi$ or $\tilde K_i^\chi$.    

\subsection {Equilibrium behavior} \label{equiferro}

Here we consider the behavior of $V^C$, $V^{C\chi}$ and ${\cal V}^\chi$ in equilibrium states above, at, and
below $T_c$. In the last case, we consider equilibrium within ergodic components,
namely in states with broken symmetry. 

\subsubsection {Limiting behaviors for $t-t_w=0$ and for $t-t_w\to \infty$} \label{static}

Before discussing the scaling properties of $V^C$, $V^{C\chi}$ and ${\cal V}^\chi$,
let us compute their limiting behaviors for small and large time differences $t-t_w$.
From the definitions~(\ref{1.6},\ref{1.72},\ref{chi2r2}) one has 
$V^C_{ij}(t,t)=V^{C,\chi}_{ij}(t,t)={\cal V}^\chi_{ij}(t,t)=0$, and the
same for the $k=0$ component. 
One can compute analytically
also the limiting values attained in equilibrium by $V^C$, $V^{C,\chi}$ and ${\cal V}^\chi$
for $t-t_w\to \infty$,
relating them to the usual static correlation function. 
Indeed,
with the definitions of Sec.~\ref{defin}, all the quantities considered are written in terms
of two-times/two-sites correlation functions. 
For large time differences these correlation functions  
can be factorized as products of one time quantities resulting in the following behavior
(details are given in Appendix IV)

\begin{eqnarray}
V^C_{ij}(\infty)&=&\lim_{t-t_w\to\infty}V^C_{ij}(t-t_w)=C_{ij,eq}(C_{ij,eq}+2m^2)
\nonumber \\
V^{C\chi}_{ij}(\infty)&=&\lim_{t-t_w\to\infty}V^{C\chi}_{ij}(t-t_w)= -m^2C_{ij,eq} \nonumber \\
{\cal V}^\chi_{ij}(\infty)&=&\lim_{t-t_w\to\infty}{\cal V}^\chi_{ij}(t-t_w)=-C_{ij,eq}^2 \label{2.10},
\end{eqnarray}
where $m$ is the equilibrium magnetization and 
$C_{ij,eq}\equiv \langle\sigma_i\sigma_j\rangle_{eq}-m^2$  
is the static correlation function.

For the $k=0$ components, from Eqs.~(\ref{2.10}) 
for $T\gtrsim T_c$, using the scaling $C_{ij,eq}\sim |i-j|^{2-d-\eta}f(|i-j|/\xi)$,
where $\xi $ is the equilibrium coherence length and $i-j$ the distance between $i$ and $j$, 
one has
\begin{equation}
V^C_{k=0}(\infty )=-{\cal V}^\chi_{k=0}(\infty)\propto \xi ^{\beta _c}
\label{statick},
\end{equation}
where
\be
\beta _c=4-d-2\eta
\ee
is an exponent related to the critical exponent $\eta $, and
\begin{equation}
V^{C\chi}_{k=0}(\infty)= 0 ,
\end{equation}
because $m=0$.
For $V^C_{k=0}(\infty)$ and ${\cal V}^\chi_{k=0}(\infty)$
the same result holds true also
below (but close to) $T_c$, since the terms containing the magnetization in Eqs.~(\ref{2.10})
can be neglected. Interestingly, the behavior
of $V^{C\chi}_{k=0}(\infty)$, on the other hand, 
is discontinuous around the critical temperature: It vanishes 
identically for $T>T_c$ while it diverges 
as $-(T_c-T)^{2\beta-\gamma}$ (where $\gamma=(2-\eta)\nu $ and $\beta$ are 
the usual critical exponents) on approaching $T_c$ from below.

\subsubsection{Scaling behavior} \label{scalingbe}

We turn now to the point we are mainly interested in, namely the scaling behavior
of $V^C$, $V^{C\chi}$ and ${\cal V}^\chi$.
To ease the notation let us introduce the symbol $V^X$, with $X=C$, $X=C\chi$, and $X=\chi$,
to denote $V^C$, $V^{C\chi}$ and ${\cal V}^\chi$, respectively.
Approaching the critical temperature the coherence length diverges and
hence a finite-size scaling analysis of the numerical data will be necessary
in Sec.~\ref{numbe}. Let us discuss here how such an analysis can be performed.  
For a finite system of linear size ${\cal L}$ we expect a scaling form 
\be
V^X_{k=0}(t-t_w)={\cal L}^{\beta _X}f_X\left (\frac{t-t_w-t_0}{{\cal L} ^{z_c}},
\frac{\xi}{{\cal L}}\right )
\label{scaleq}
\ee
where $t_0$ is a microscopic time, $z_c$ is the dynamic critical exponent, and $f_X(x,y)$ 
a scaling function (in the following, in order to simplify
the notation, we will always denote scaling functions with
an $f$, even if, in different cases, they may have different functional forms). 
Away from the critical point,
matching the large $t-t_w$ behavior
of Eq.~(\ref{scaleq}) 
with the large time difference limits $V^C_{\infty}$, ${\cal V}^\chi_{\infty}$ 
of Eq.~(\ref{statick}) implies 
$f_C(x,y)\sim f_\chi(x,y)\sim \xi ^{\beta_c}/{\cal L} ^{\beta _X}$.
Since only the ratio $\xi /{\cal L}$ must enter $f_X$
this fixes the exponents $\beta_C=\beta_\chi=\beta _c$.
Finally, Eq.~(\ref{quadrato}) implies that also $\beta_{C\chi}$
takes the same value and, in conclusion
\be
\beta _X\equiv \beta _c
\ee
for all the quantities.
Letting ${\cal L}^{z_c}= (t-t_w-t_0)$ 
in Eq.~(\ref{scaleq}) implies 
\be
V^X_{k=0}(t-t_w)= (t-t_w-t_0)^{b_c}
f_X\left (\frac{t-t_w-t_0}{\xi ^{z_c}}\right ),
\label{s1}
\ee
where $f_X\left [(t-t_w-t_0)/\xi ^{z_c}\right ]$ is a shorthand for
$f_X\left [1,(t-t_w-t_0)/\xi ^{z_c}\right ]$,
and $b_c=\beta _c/z_c$. Assuming that there is no
dependence on $\xi $ for small time differences $t-t_w$ leads to
$f_X(x)\sim const$ in this regime. This implies
\be
V^X_{k=0}(t-t_w)\sim  (t-t_w-t_0)^{b_c}
\label{stb}
\ee 
for $(t-t_w-t_0)\ll \xi ^{z_c}$.

\subsubsection{Numerical studies} \label{numbe}

In this Section we study numerically the equilibrium behavior of 
the two-dimensional Ising model, where $z_c\simeq 2.16$ and $b_c\simeq 0.69$,
and check the scaling laws derived above.

Before presenting the results let us comment on the method used to
compute the $k=0$ components.
For $T\neq T_c$, for any 
$t$ and $t_w$, $V^X_{ij}(t-t_w)$ decay over a distance $i-j$ at most of order of 
$\xi $. Then,
performing the sum in Eq.~(\ref{keq0}) 
over the whole system one introduces a number of order 
$[({\cal L}-\xi )/\xi]^d$ of terms whose average value is negligible.
However, due to the limited statistics of the simulations,
such terms are not efficiently averaged and introduce noisy contributions
which, with the definition~(\ref{keq0}) sum up to produce an overall 
noise of order $[({\cal L}-\xi )/\xi]^{d/2}$.
For ${\cal L}$ much larger than $\xi $ this quantity is large and lowers the numerical accuracy.
Therefore, since one knows that the average of that noise is zero,
the most efficient way of computing $V^X_{k=0}$ is
to sum only up to distances $i-j =l\gtrsim \xi $. We have checked that the two procedures
(namely summing over all the sites $i,j$ of the system or restricting to those
with $i-j \le l$) give the same results within the numerical uncertainty.    
We anticipate that in the study of non-equilibrium after a quench
below $T_c$ presented in Sec.~\ref{beque}, similar considerations
apply with $\xi $ replaced with $L(t)$, the typical size of domains.
Clearly, at $T_c$ where $\xi =\infty$ such a procedure cannot be applied and
the sum must be performed over the whole system. 

\begin{figure}
    \centering
   \rotatebox{0}{\resizebox{.45\textwidth}{!}{\includegraphics{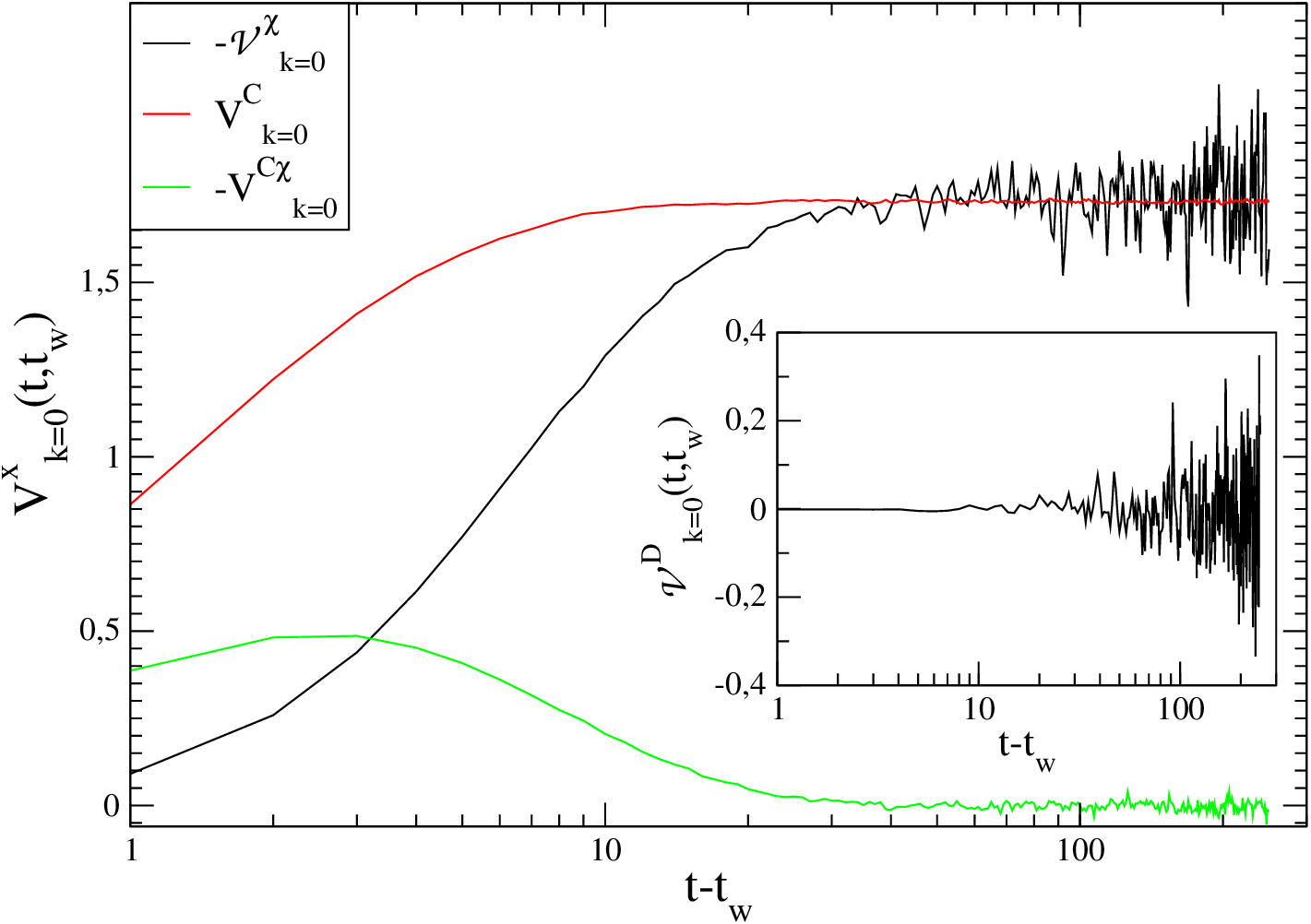}}}
   \rotatebox{0}{\resizebox{.45\textwidth}{!}{\includegraphics{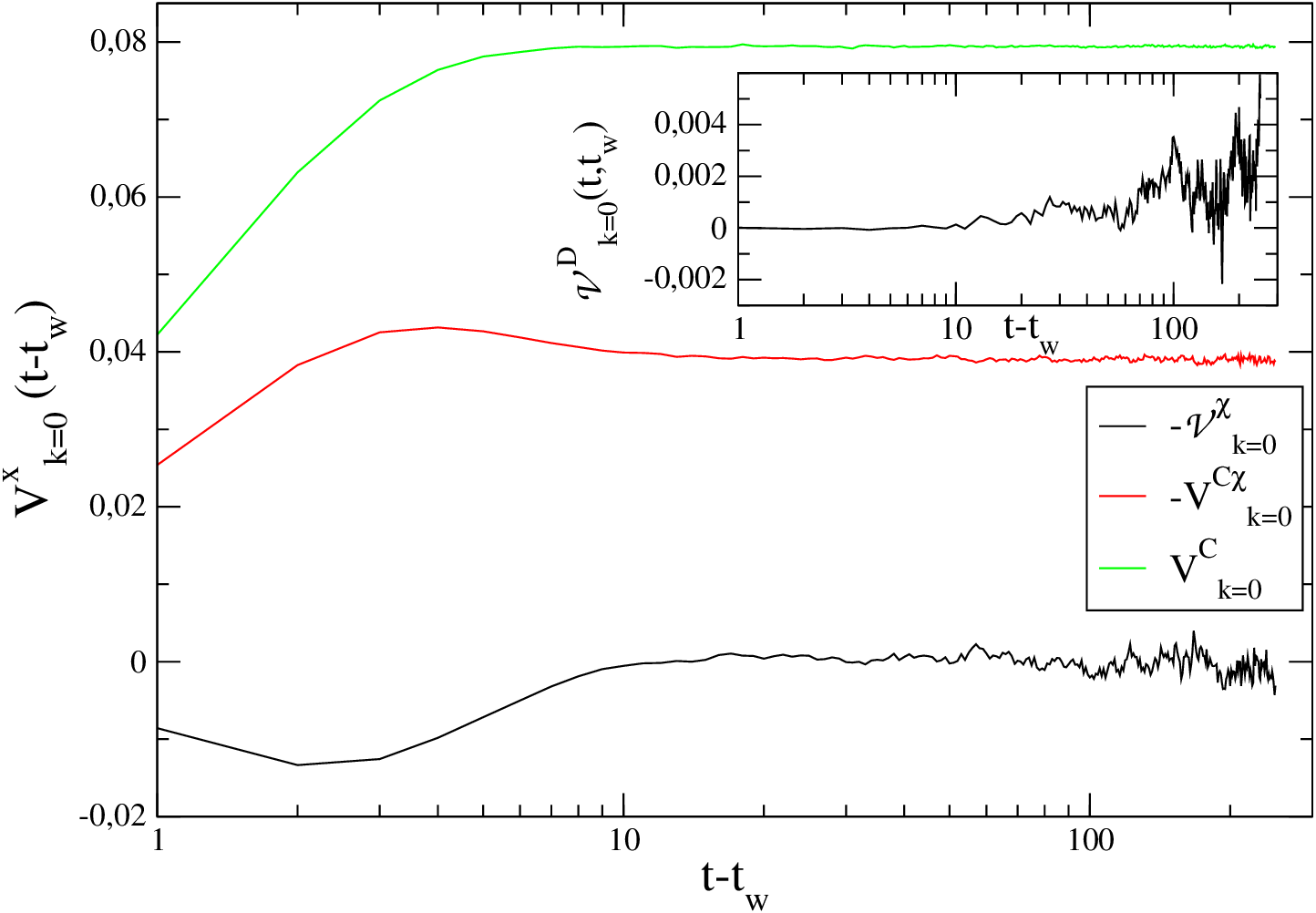}}}
   \vspace{0.5cm}
    \caption{(Color online). $V^C_{k=0}(t-t_w)$, $-{\cal V}^{\chi}_{k=0}(t-t_w)$
    and $-V^{C\chi}_{k=0}(t-t_w)$ are plotted against $t-t_w$ in equilibrium
    conditions at $T=3.5 >T_c$ 
    (left panel), where
    $\xi \simeq 1.98$ and at $T=1.5<T_c$ (right panel) 
    where $\xi \simeq 0.88$. In the insets ${\cal V}^D_{k=0}(t,t_w)$
    is plotted against $t-t_w$. The system size is
    ${\cal L}=10^3$ and $l=10^2$.}
\vspace{1cm}
\label{d2eqabovebelow}
\end{figure}

Starting from the case $T>T_c$, in the left part of Fig.~\ref{d2eqabovebelow} 
we plot $V^C$, $V^{C,\chi}$ and ${\cal V}^\chi$ as functions of $t-t_w$.
$V^C_{k=0}$ and $-{\cal V}^\chi _{k=0}$ grow monotonically
to the same limit~(\ref{statick}), while $V^{C\chi}_{k=0}$
has a non monotonic behavior vanishing for large time differences.
In the inset, by plotting ${\cal V}^D_{k=0}(t,t_w)$
(we recall that $V^C_{k=0}(t,t)\equiv 0$ for Ising spins)
we confirm the SOFDT~(\ref{quadrato}).

In the case of quenches below $T_c$ (right part of Fig.~\ref{d2eqabovebelow}),
we obtained the broken symmetry equilibrium state by preparing an ordered
state (i.e. all spins up) and then letting it relax at the working temperature
to the stationary state. In this case
the behavior of $V^C$, $V^{C,\chi}$ and ${\cal V}^\chi$ is similar to the case $T>T_c$, with the difference that also
${\cal V}^\chi _{k=0}$ has a non monotonic behavior. 
We have checked that for temperatures close to $T_c$, both above and below $T_c$
(i.e. for $T=2.28$ and $T=2.25$) $V^C$, $V^{C,\chi}$ and ${\cal V}^\chi$ grow as $(t-t_w-t_0)^{b_X}$
with $b_X$ consistent with the expected value $b_c$, 
as expressed in Eq.~(\ref{stb}). 

\begin{figure}
    \centering
    
   \rotatebox{0}{\resizebox{.45\textwidth}{!}{\includegraphics{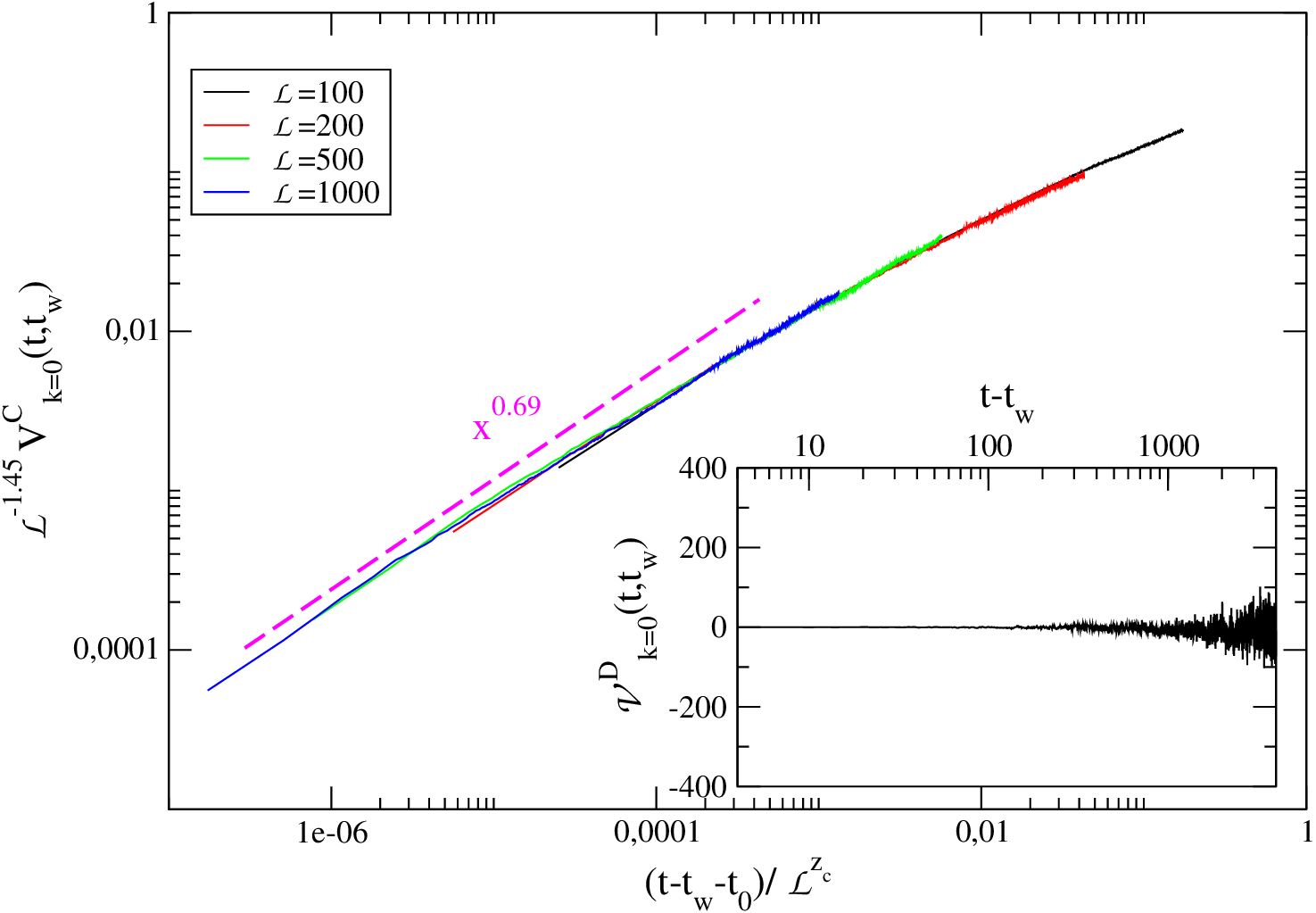}}}
    \caption{(Color online). ${\cal L}^{-1.45} V^C_{k=0}(t-t_w)$,  is plotted 
    against $(t-t_w-t_0)/{\cal L}^{z_c}$ (with $t_0=0.475$), in equilibrium
    conditions at $T_c$ for different values of ${\cal L}$. 
    The dashed line is the expected power-law behavior 
    $[(t-t_w-t_0)/{\cal L}^{z_c}]^{b_c}$ ($b_c=0.69$).
    In the inset ${\cal V}^D_{k=0}(t,t_w)$ is plotted 
    against $t-t_w$ for ${\cal L}=10^3$.}
\label{d2eqat}
\end{figure}

In order to study the critical behavior we have equilibrated the system at $T_c$ using
the Wolff cluster algorithm~\cite{Wolff}. 
In Fig.~\ref{d2eqat} we present a finite size scaling analysis of the data.
In view of Eq.~(\ref{scaleq}) we plot  ${\cal L}^{-1.45} V^C_{k=0}$
for different ${\cal L}$ against $(t-t_w-t_0)/{\cal L}^{z_c}$,
where $t_0= 0.475$ and the exponent $1.45$ (in good agreement with the
expected value $\beta_c= 1.5$) have been obtained by requiring the best data collapse.  
All the data exhibit a nice collapse on a
unique master-curve. The master-curve grows initially as a power law with an
exponent $0.69$ in good agreement with $b_c$, as expected
from Eq.~(\ref{stb}), and than tends toward saturation for 
$t-t_w+t_0\gg {\cal L}^{z_c}$. A similar behaviour is observed for the other $V^C$, $V^{C\chi}$,
apart from the sign, since ${\cal V}^\chi_{k=0}$ and $V^{C\chi}_{k=0}$ are negative for
large $t-t_w$.
 
\subsection{Non equilibrium} \label{quench}

\subsubsection{Critical quench} \label{critque}

In this Section we consider a ferromagnetic system quenched from an equilibrium
state at infinite temperature to $T_c$. Numerical results are presented
for $d=2$. In $d=3$ the situation is qualitatively similar although our data are 
too noisy to extract precise quantitative information.
With $T=T_c$ we expect a scaling form as in Eq.~(\ref{scaleq}) where
the role played by ${\cal L}$ is now assumed by $L(t_w)\sim t_w^{1/z_c}$.
Letting $t-t_w\gg t_0$, one has 
\be
V^X_{k=0}(t,t_w)\simeq t_w^{b_c}f_X\left ( \frac{t}{t_w}\right ),
\label{scaltcc}
\ee
where $f_X\left (t/t_w\right )$ is a shorthand for $f_X\left (t/t_w,\infty \right )$,
and the short time behavior~(\ref{stb}).
Notice that this scaling, together with the equilibrium one~(\ref{s1}), are
consistent with the results of Ref.~\cite{annibale}
where the same forms are obtained with $b_c=(4-d-2\eta )/z_c=(4-d)/2$,
since in the spherical model $\eta= 0$ and $z_c=2$.  

\begin{figure}
    \centering
    
   \rotatebox{0}{\resizebox{.45\textwidth}{!}{\includegraphics{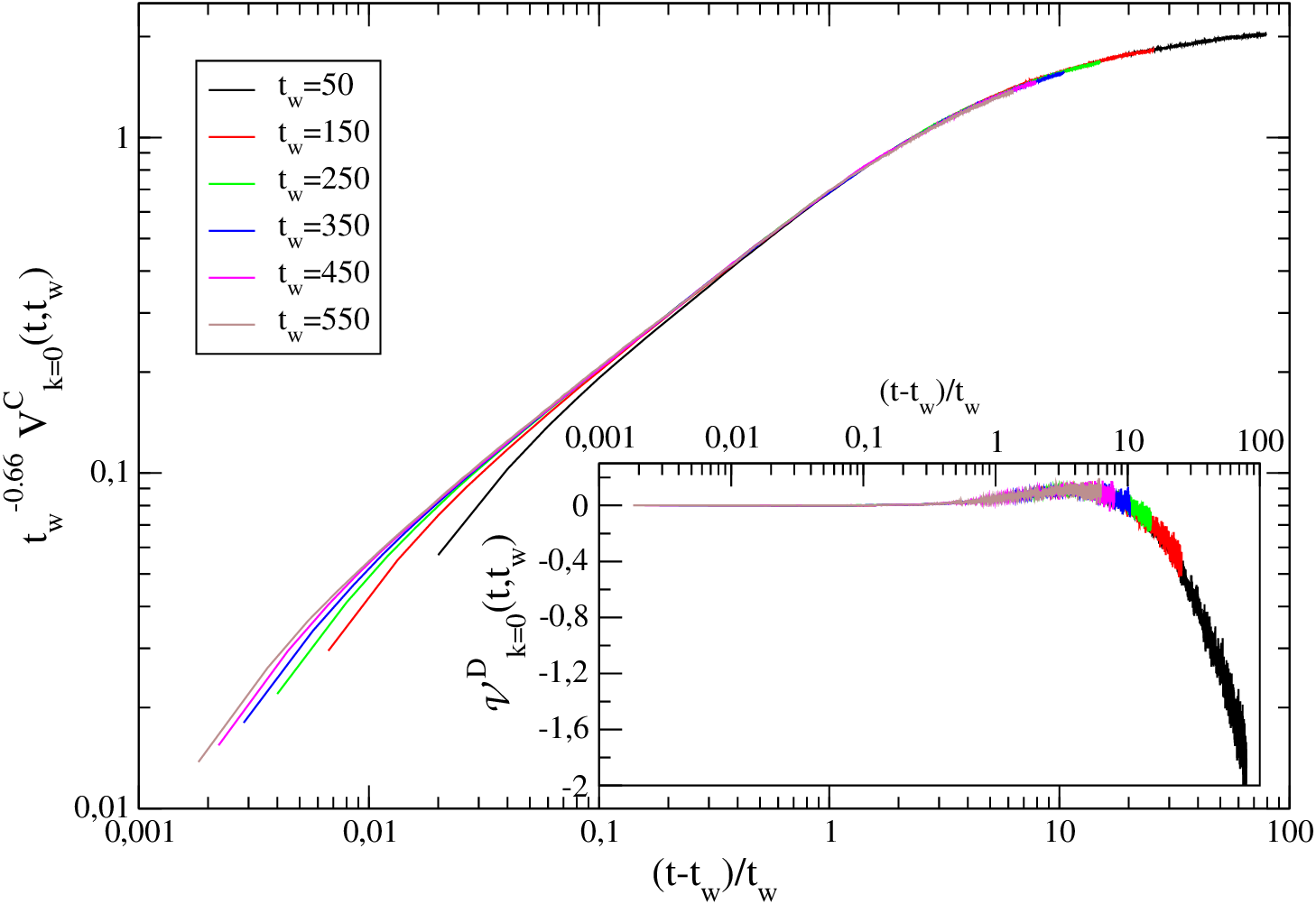}}}
   \rotatebox{0}{\resizebox{.45\textwidth}{!}{\includegraphics{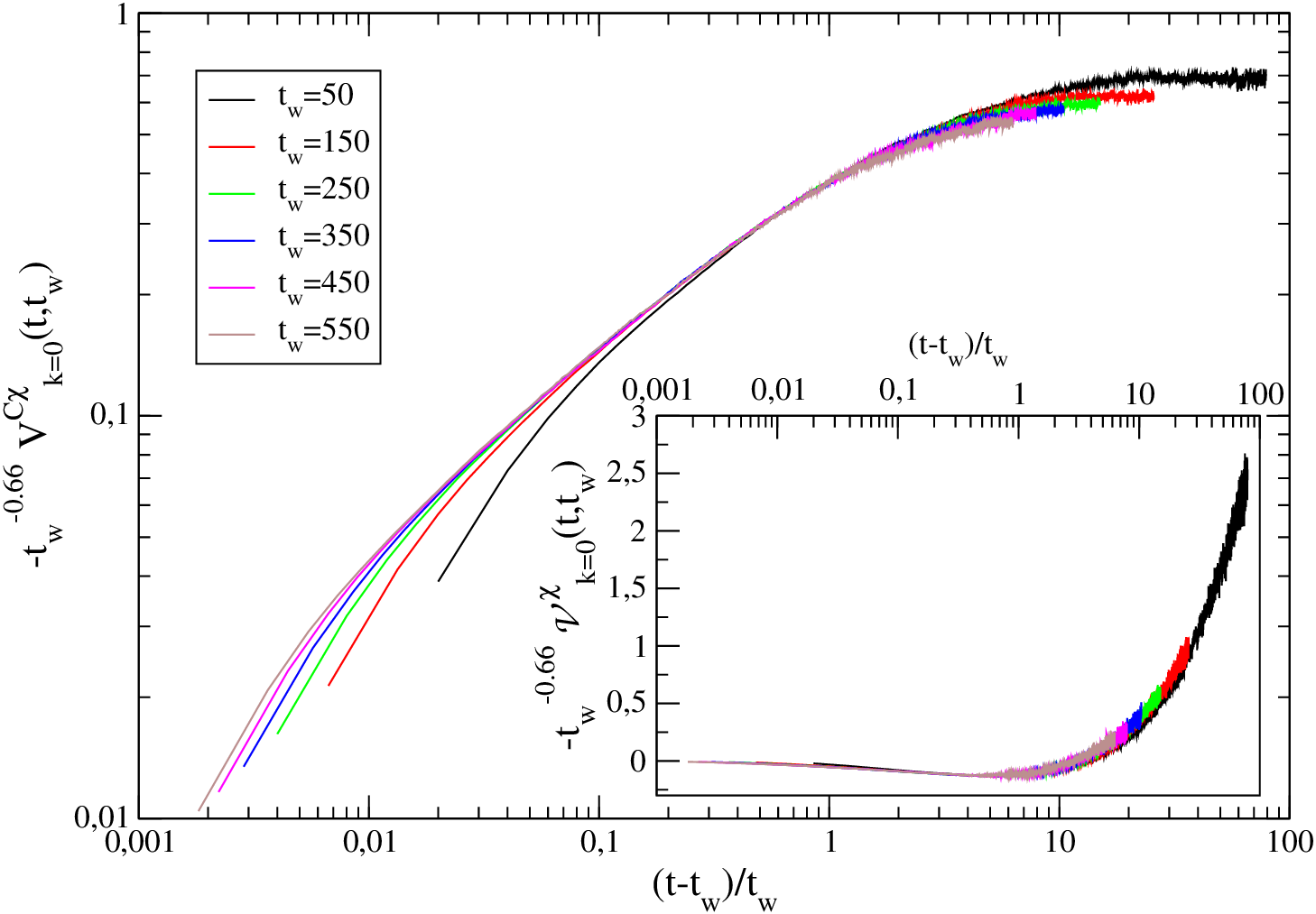}}}
   \vspace{.5cm}
    \caption{(Color online). $t_w^{-0.66}V^C_{k=0}(t-t_w)$ (left panel, log-log scale), 
    $-t_w^{-0.66}V^{C\chi}_{k=0}(t-t_w)$ (right panel, log-log scale)
    and $-t_w^{-0.66}  {\cal V}^{\chi}_{k=0}(t-t_w)$ (inset of right panel, log-linear scale) 
    are plotted against $(t-t_w)/t_w$ for a quench to $T_c$ in $d=2$. In the inset of the left panel 
    ${\cal V}^D_{k=0}(t,t_w)$ is plotted against $(t-t_w)/t_w$.}
\label{d2neqat}
\end{figure}

The behavior of $V^C$, $V^{C,\chi}$ and ${\cal V}^\chi$ is shown in Fig.~\ref{d2neqat}. 
By plotting $t_w^{-0.66} V^X_{k=0}(t,t_w)$ vs $(t-t_w)/t_w$
one observes a good collapse of the curves for $(t-t_w)/t_w$ sufficiently large.
Lack of collapse for $t-t_w\lesssim t_0$ is expected due to the $t_0$-dependence
in the scaling form (\ref{scaleq}) and those derived from it.  
The exponent $0.66$ is in good agreement with the expected one $b_c\simeq 0.69$, 
and this confirms that the scaling~(\ref{scaltcc}) is obeyed. 
In the short time difference regime, for $t-t_w\ll t_w$, these quantities behave as in equilibrium,
and in particular the relation~(\ref{quadrato}) is obeyed, as it is shown in the inset of 
the left panel of Fig.~\ref{d2neqat}. 
For $t-t_w\gtrsim t_w$ the relation~(\ref{quadrato})
breaks down and the asymptotic regime is entered. In this time domain
$V^C$ and $V^{C\chi}$ approach constant values in the large $t$ limit. 
For $V^C_{k=0}$ this can be understood as follows:
Writing the sum~(\ref{keq0}) as an integral 
\be
V^C_{k=0}(t,t_w)=\int d {\bf r} \,\, V^C(r,t,t_w)
\ee
where $r=\vert i-j\vert$, and invoking the clustering property, by
factorizing  
$V^C(r,t,t_w)$ for $t\to \infty$, one has
\be
V^C_{k=0}(t,t_w)
\simeq \int d{\bf r} \,\, C_r(t_w)C_r(t).
\label{u}
\ee
Using the scaling of the correlation function $C_r(t)=t^{-(d-2+\eta)/z_c}f(r/t^{1/z_c})$, 
with the small $x$ behavior $f(x)\sim x^{-(d-2+\eta)}$, 
Eq.~(\ref{u}) becomes
\be
\lim _{t\to \infty} V^C_{k=0}(t,t_w)= ct_w^{b_c}  
\label{appr}
\ee
with $c=\int d^dx x^{-(d-2+\eta)}f(x)$.
Notice that the asymptotic value~(\ref{appr}) approached by
$V^C_{k=0}$ (and $V^{C\chi}_{k=0}$)
are increasing functions of $t_w$. 
This mechanism makes
$\lim _{t_w\to \infty }\lim _{t\to \infty}V^C_{k=0}(t,t_w)=\infty$, and in
this sense the limit~(\ref{statick}) is recovered, bearing in mind that $\xi =\infty$.
Moreover, $\lim _{t_w\to \infty }\lim _{t\to \infty}V^{C\chi}_{k=0}(t,t_w)/V^{C}_{k=0}(t,t_w)$, 
is a $t_w$-independent constant as was found in~\cite{corbandolo}.
We also observe that $V^{C\chi}_{k=0}$ going
to a constant value is a different behavior with respect to
the spherical model~\cite{annibale}, where this quantity vanishes
for $t\to \infty$.
The quantity ${\cal V}^\chi _{k=0} $ has a different behavior, 
in that it diverges for $t\to \infty $ for any value of $t_w$.
Therefore, at variance with $V^C_{k=0}$ and $V^{C\chi}_{k=0}$, 
the limit~(\ref{statick}) is always recovered,
irrespectively of $t_w$. 
This is a general property of susceptibilities.
Considering the linear case for simplicity, from Eq.~(\ref{abcde}) one
sees that $\chi _i$ can be written as an average of a one-time quantity over a process
where the Hamiltonian is changed at $t_w$. Since
the average of a one-time quantity must tend to its (perturbed) equilibrium value
for large $t$ (even if the Hamiltonian has been modified at $t_w$),
this explains why $\lim _{t\to \infty}\chi _i(t,t_w)$ is independent of $t_w$.
An analogous argument holds for ${\cal V}^\chi _{ij}$. Indeed, recalling Eqs.(\ref{corchih},\ref{r222}),
for $i\neq j$ also this quantity can be written as an average of a one-time quantity.
The same property holds for the equal site contribution since, 
according to Eq. (\ref{hyp}), it is  
${\cal V}^\chi _{ii}(t,t_w)=-\chi _i^2(t,t_w)$.
The limit~(\ref{statick}) is then satisfied irrespectively of $t_w$.

\subsubsection{Quench below $T_c$} \label{beque}

In this Section we consider a ferromagnetic system quenched from an equilibrium
state at infinite temperature to $T<T_c$, in $d=1,2$.

Let us recall the behavior of $C$ and $\chi $ in a quench from
$T=\infty$ to $T<T_c$. 
In the large-$t_w$ limit $C$ obeys the following additive 
structure~\cite{addit}
\be
C(t,t_w)\simeq C_{st}(t-t_w)+C_{ag}(t,t_w).
\label{additive}
\ee
Here $C_{st}$ is the contribution provided by bulk spins which are
in local equilibrium. 
This term vanishes for quenches to $T=0$.
$C_{ag}(t,t_w)$ is the aging contribution originated by the presence of interfaces
which scales as
\be
C_{ag}(t,t_w)=t_w^{-b}f\left(\frac{t}{t_w}\right),
\label{agingc}
\ee
with $b=0$ and the property~\cite{noiteff}  
\be
f(x)\sim x^{-\lambda}
\ee
for large $x$, where $\lambda $ is related to the Fisher-Huse exponent.

A decomposition analogous to Eq.~(\ref{additive}) holds for $\chi (t,t_w)$,
with $\chi _{st}(t-t_w)=\chi _{eq}(t-t_w)=C_{eq}(t,t)-C_{eq}(t,t_w)$
and 
\be
\chi_{ag}(t,t_w)=t_w^{-a}g\left (\frac{t}{t_w}\right),
\label{agingchi}
\ee
with $g$ behaving as~\cite{noiteff} 
\be
g(x)\sim x^{-a}
\ee
for large $x$. The exponent $a$ depends on spatial dimensionality 
so that $a>0$ for $d>d_l$, $d_l$ being the lower critical dimensionality, and $a=0$
for quenches at $d=d_l$ with $T=0$~\cite{noiteff,rough,graphs,expa}. 
Notice that, at variance with the critical quench, the non-equilibrium 
exponents are not related to the equilibrium ones, since the additive 
form~(\ref{additive}) splits them in separate terms.

The addictive structure~(\ref{additive}), which is expected also
for the $V^X$, opens the problem of
disentangling the stationary from the aging contributions to allow the
separate analysis of their scaling properties. In order to do this one usually
enforces the knowledge of the time-sectors where the stationary and
the aging terms contribute significantly.
Specifically, working in the short time difference regime, namely
with $t_w\to \infty$ and $t-t_w$ finite, the aging term is 
constant and one can study the behavior of the stationary one.
On the contrary, in the aging regime with $t_w \to \infty$ and
$t/t_w$ finite, $C_{st}(t-t_w)\simeq 0$ and one has direct access to $C_{ag}$.
The same procedure can be applied to isolate the stationary and
aging contributions to the $V^X$, as will be done in Section~\ref{beque}.
However, in so doing one effectively separates the two contributions only
in the limit $t_w\to \infty$. In numerical simulations, where finite values
of $t_w$ are used, a certain mixing of the two is unavoidable and may affect the
results.
Furthermore, this technique fails in systems where 
(in contrast to the ferromagnetic model
considered here) we do not have a precise knowledge of the time sectors where
stationary and aging terms contribute.

For the $V^X$
a more elegant and effective technique to isolate the aging from
the stationary terms relies on the SOFDT. Indeed, according to Eq.~(\ref{quadratoa}),
an exact cancellation occurs in ${\cal V}^D$ 
between the stationary (equilibrium) terms, so that only the aging
behavior is reflected by ${\cal V}^D$. In other words, recalling
the discussion at the end of Section~\ref{ineqj}, the quantity
$\widehat D$ does not produce any correlation in equilibrium and hence
what is left in ${\cal V}^D$ are the correlations due to aging.
This fact will be enforced in Section~\ref{beque}.

Let us now consider the behavior of $V^C$, $V^{C,\chi}$ and ${\cal V}^\chi$ in $d=1,2$.

\vspace{.5cm}
{\bf d=1, quench to T=0}
\vspace{.5cm}

\begin{figure}
    \centering
    
   \rotatebox{0}{\resizebox{.45\textwidth}{!}{\includegraphics{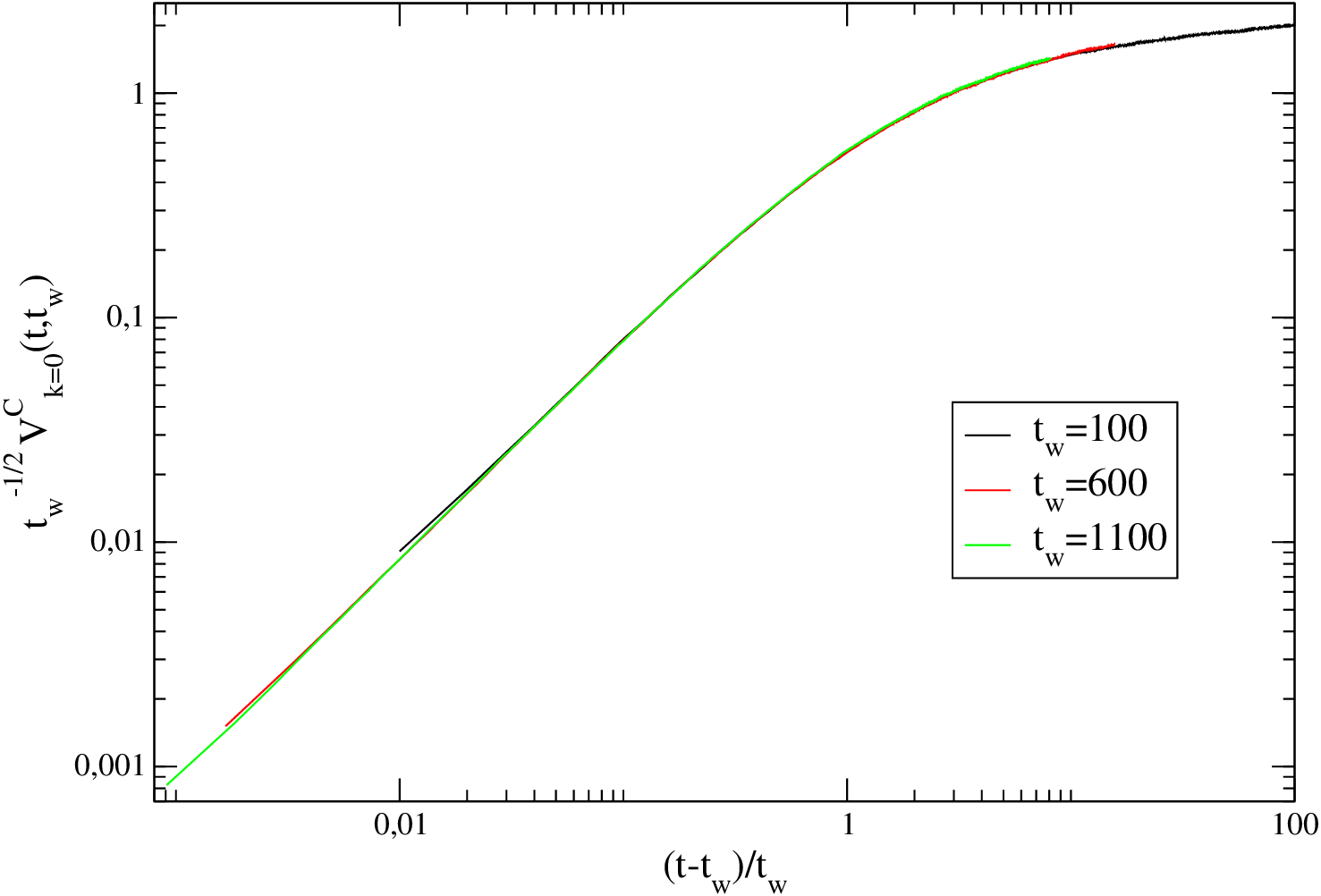}}}
   \rotatebox{0}{\resizebox{.45\textwidth}{!}{\includegraphics{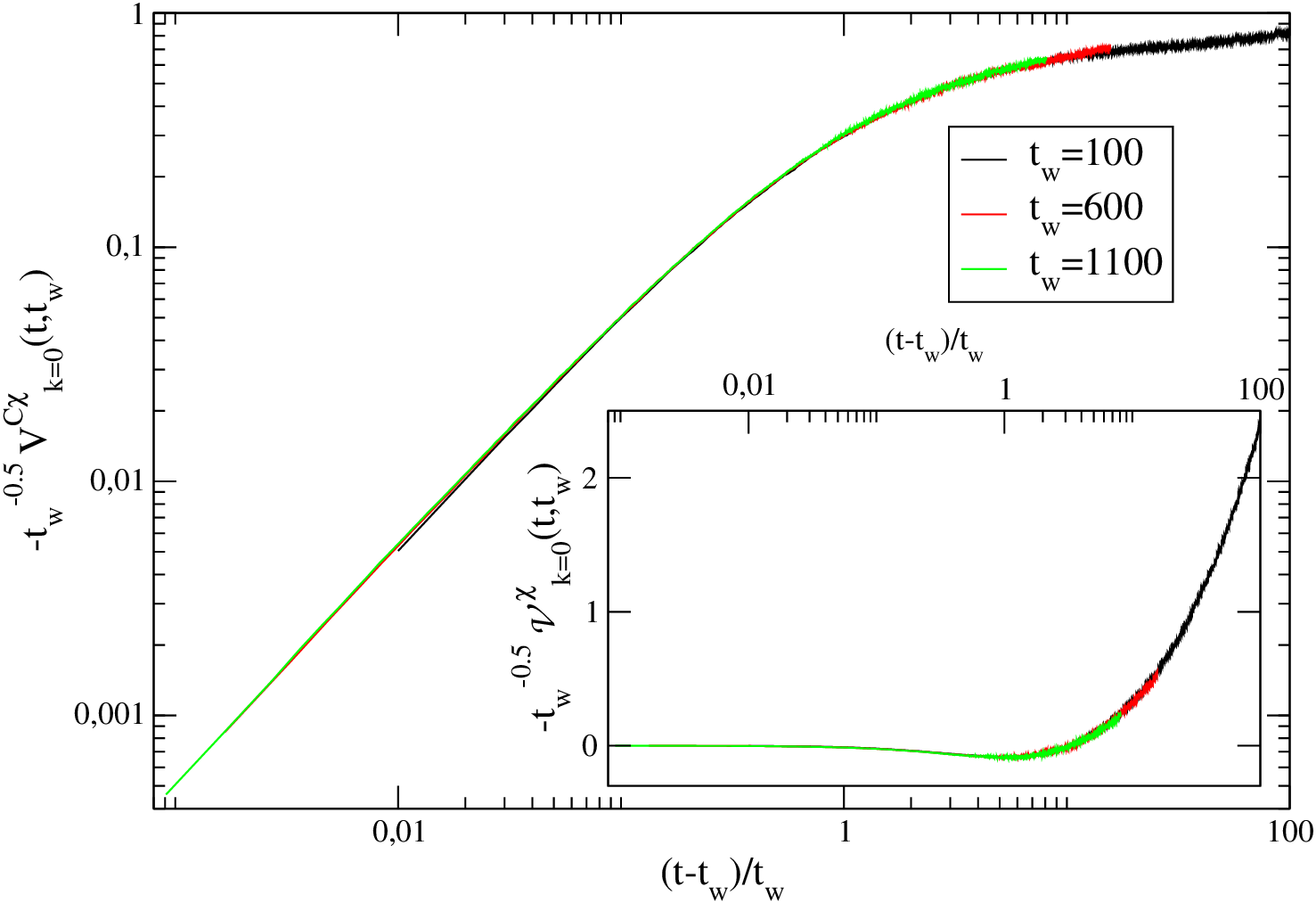}}}
\vspace{.5cm}    
\caption{(Color online).
   $t_w^{-1/2}V^C_{k=0}(t-t_w)$ (left panel, log-log scale), 
    $-t_w^{-1/2}  V^{C\chi}_{k=0}(t-t_w)$
    (right panel, log-log scale) and $-t_w^{-1/2}V^{\chi}_{k=0}(t-t_w)$ (inset of right panel,
     log-linear scale) 
     are plotted against $(t-t_w)/t_w$ for different values of $t_w$ in the key in a quench to $T=0$
    in $d=1$.}  
\vspace{1cm}
\label{d1neqbelow}
\end{figure}

As explained in~\cite{noiteff} the dynamical features of a 
system at the lower critical dimension quenched to $T=0$
are those of a quench into the ordered region, rather than those
of a critical quench, due to a non-vanishing 
Edwards-Anderson order parameter $q_{EA}=\lim _{t\to \infty} \lim _{t_w\to \infty}C(t,t_w)$.
Since at $T=0$ there are no stationary contributions we expect 
$V^X_{k=0}(t,t_w)=V^X_{k=0,ag}(t,t_w)$, with the scaling
\be
V^X_{k=0,ag}(t,t_w)=t_w^{a_X}f_X\left(\frac{t}{t_w}\right).
\label{scalbelowx}
\ee
The behavior of these quantities is shown in 
Fig.~\ref{d1neqbelow}.
By plotting $t_w^{-1/2} V^X(t,t_w)$ vs $t/t_w$
one observes an excellent collapse of the curves
(tiny deviations from the mastercurve for small values of $t-t_w$
are due to the $t_0$-dependence, as discussed above).
This implies that Eq.~(\ref{scalbelowx}) is obeyed with
$a_X=1/2$. Notice that, 
for large values of $t/t_w$, $V^C$ and $V^{C\chi}$ seem to approach
constant values whereas 
${\cal V}^\chi_{k=0}$ grows as ${\cal V}^\chi\propto t^{1/2}$.
Then one has the limiting behavior $f_X(x)\sim x^{\lambda _X}$,
with values of the exponents consistent with $\lambda _C=\lambda _{C\chi}=0$ and $\lambda _\chi=1/2$.

\vspace{.5cm}
{\bf d=2, quench to $\mathbf{0<T<T_c}$}
\vspace{.5cm}

\begin{figure}
    \centering
    
   \rotatebox{0}{\resizebox{.45\textwidth}{!}{\includegraphics{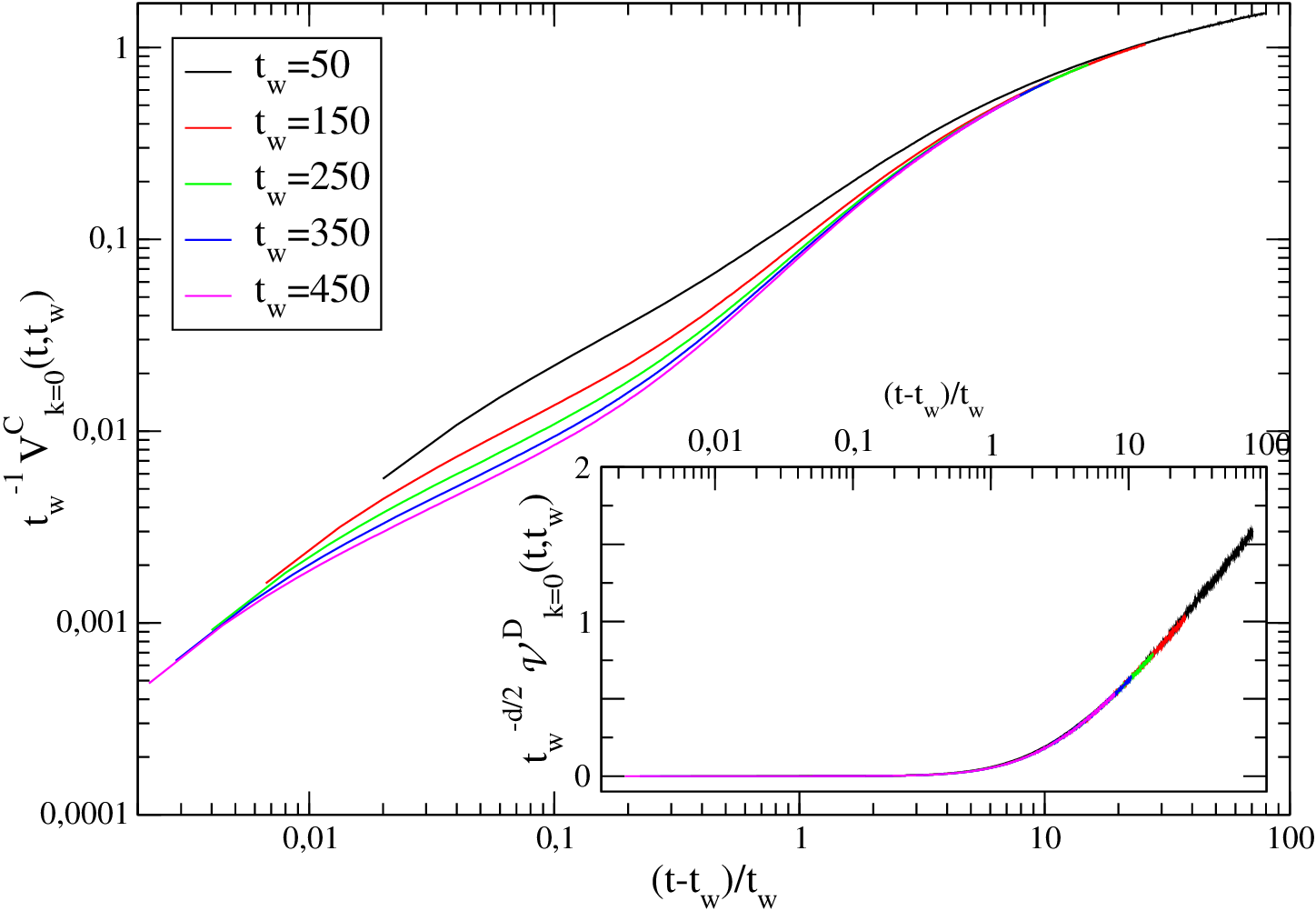}}}
   \rotatebox{0}{\resizebox{.45\textwidth}{!}{\includegraphics{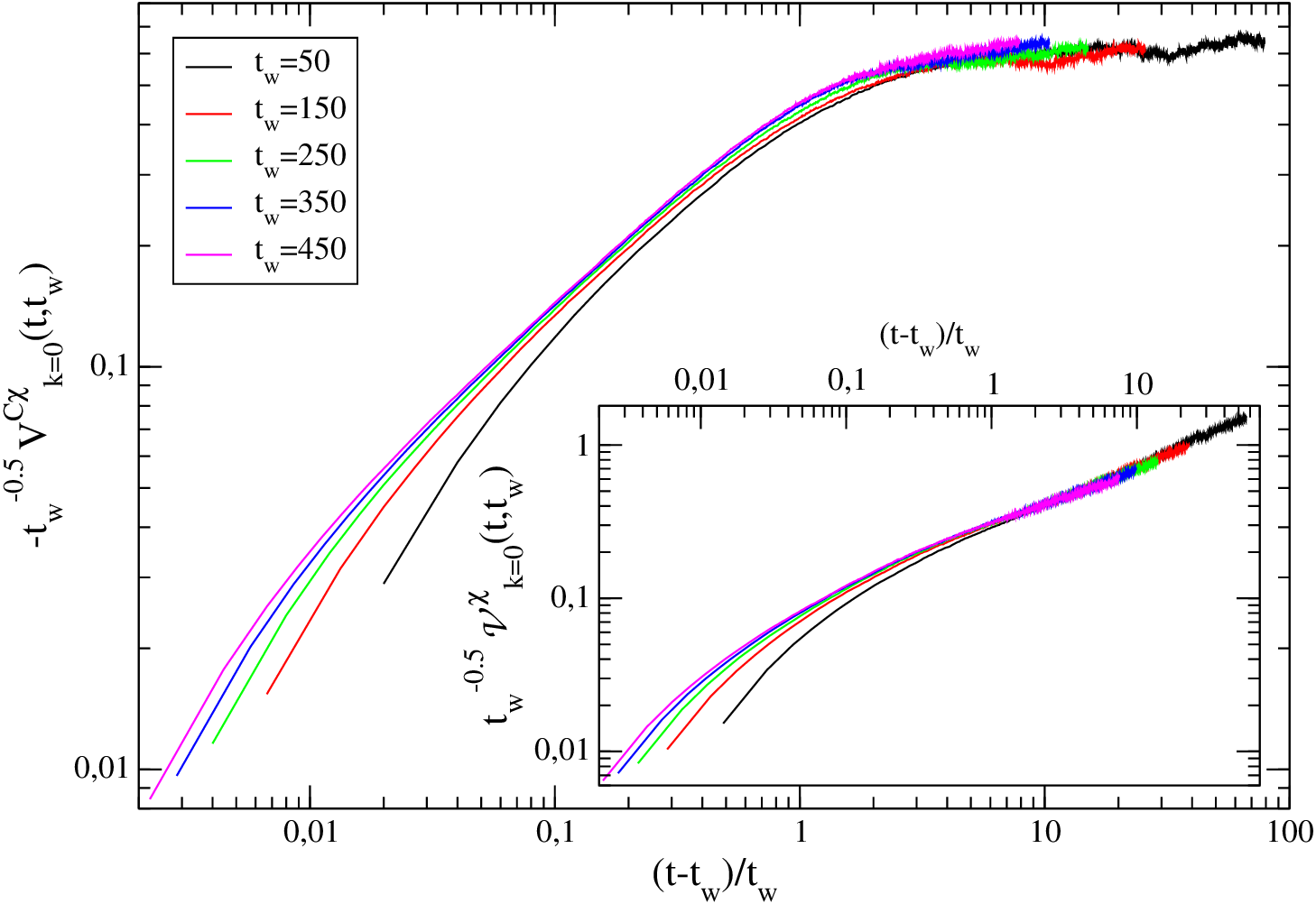}}}
\vspace{.5cm}    
    \caption{(Color online).
    $t_w^{-1}V^C_{k=0}(t-t_w)$ (left panel), 
    $-t_w^{-1/2}  V^{C\chi}_{k=0}(t-t_w)$
    (right panel) and $t_w^{-1/2}{\cal V}^\chi_{k=0}(t-t_w)$ (inset of right panel) 
     are plotted against $(t-t_w)/t_w$ for different values of $t_w$ in the key in a quench to $T=1.5$
    in $d=2$.  
    In the inset of the left panel ${\cal V}^D_{k=0}(t,t_w)$ is plotted against $(t-t_w)/t_w$.}  
\label{d2neqbelow}
\end{figure}

In this case we consider quenches to finite temperatures, and hence stationary
contributions are present. We expect that one can select between the stationary
and the aging contributions to $V^X(t,t_w)$ by considering
the short time limit and the aging regimes separately.
In the former case, only the stationary terms contribute
and then we expect the relation~(\ref{quadrato}) to
be obeyed. This is shown in the inset of the left panel of Fig.~\ref{d2neqbelow}
where the relation~(\ref{quadrato}) is observed to hold for $t-t_w\lesssim t_w$.  
In the aging regime one selects the aging contribution scaling as
in Eq.~(\ref{scalbelowx}).
Indeed, by plotting in Fig.~\ref{d2neqbelow} $t_w^{-a_X} V^X(t,t_w)$ vs $(t-t_w)/t_w$
one observes an asymptotic collapse of the curves with $a_C=1$ and $a_{C\chi}=a_\chi=1/2$.
A residual
$t_w$ dependence can be observed (particularly for $V^C$) 
that tends to reduce on increasing $t_w$.
This suggests interpreting this corrections as being produced by the stationary contributions
which, due to the limited values of $t_w$ used in the simulations, are not yet completely
negligible. A clear confirmation of this interpretation comes from the inspection
of the behavior of $V^D_{k=0}$ in the inset of the left panel of Fig.~\ref{d2neqbelow}.
Indeed one observes that, at variance with $V^C_{k=0}$, this quantity
exhibits an excellent scaling for every value of $t_w$, due to the fact that the
stationary contributions do not contribute to  $V^D_{k=0}$.
This suggests the use of $V^D_{k=0}$ to study aging behaviors in
more complex non-equilibrium systems, such as spin glasses, where
the nature of the stationary contribution has not yet been clarified.

The results for $V^C$ in $d=1$ and $d=2$ suggest that the scaling exponent 
depends on space dimension as $a_C=d/2$.
This can be understood on the basis of an
argument which, for simplicity, is presented below for the case $d=1$.
Let us consider an interface $I$ at position $x_I(t_w)$ at time $t_w$.
Suppose that at time $t$, 
$I$ has moved to a new position $x_I(t)>x_I(t_w)$.
To start with, let us suppose that $I$ is the only interface
present in the system and that $t-t_w\ll t_w$. Let us indicate by ${\cal R}_I$
the region where $\widehat C_i(t,t_w)=-1$ (in the present case 
the region $x(t_w)<i<x(t)$ swept out by the interface).
Since $V^C_{ij}$ is the correlation functions of the $C_i$'s and the
quench is effectively made to below $T_c$, $V^C_{k=0}$
is proportional to the volume $V({\cal R}_I)=x_I(t)-x_I(t_w)$ of the region
${\cal R}_I$. For $t-t_w \ll t_w$ interfaces can be considered as yielding independent
contributions and the above argument can be extended to the physical case with many
interfaces. In doing that one simply has to replace $V({\cal R}_I)$ 
with its typical value $V({\cal R})$ obtained by averaging over the behavior of all
the interfaces.
Since the typical value of $V({\cal R})$ is $L(t)-L(t_w)$
one obtains $V^C_{k=0}(t,t_w)\propto L(t)-L(t_w)$. Repeating the argument for
generic $d$ one finds $V^C_{k=0}(t,t_w)\propto L^d(t)-L^d(t_w)$. 
For $t-t_w\gtrsim t_w$ the situation is more complex because 
in this time domain another interface $J$ may move
into the region swept out by the interface $I$ and one cannot disentangle
their contributions. The situation simplifies again in the
limit $t-t_w \to \infty$, because $x_I(t)\gg x_I(t_w)$ (we assume,
without loss of generality, that $I$ has moved in the direction of increasing $i$). 
In this case, in the region
swept ut by the interface, the configuration
of the system at $t_w$ was characterized by many domains of
different sign. For an interface separating positive spins on the left of it from
positive ones, $\widehat C_i(t,t_w)$ is equal to the sign of the domain 
to which the $i$-th spin belonged at $t_w$. Then, almost all the contributions to 
$V^C_{k=0}$ cancels, because of these alternating signs. The only unbalance between
positive and negative contributions comes from the region around $x_I(t)$.
Indeed, the interface can build up a positive contribution to $V^C_{k=0}$
if it is not centered on the middle of the domain located there at $t_w$.
This contribution is of order $L(t_w)$
($V({\cal R})\sim L^d(t_w)$, for generic $d$) leading to the saturation
of $V^C_{k=0}$ to a $t_w$-dependent value for large $t$. 
In conclusion, from the argument above we obtain, in both of the regimes
$t-t_w\ll t_w$ and $t-t_w\gg t_w$, a behavior consistent with
Eq.~(\ref{scalbelowx}) with $a_C=d/2$ and $f_C(x)\simeq (x^{d/2}-1)$ for 
$t-t_w\ll t_w$ and $\lim _{x\to \infty}f_C(x)=const$.
The same result is found in the soluble large-N model~\cite{nontris}.
Another way to understand the behavior of $V^C$ is the following:
Factorizing for $t-t_w\to \infty $ as $V^C_{k=0}
=\int d{\bf r} 
\langle \sigma _i(t)\sigma _j(t)\rangle \langle \sigma _i(t_w)\sigma _j(t_w)\rangle $, 
using the scaling $\langle \sigma _i(t)\sigma _j(t)\rangle =
g(r/L(t))$ and performing the integral one has 
\be
\int d^dr g(r/L(t))g(r/L(t_w))=L(t_w)^d\int d^dx g(xL(t_w)/L(t))g(x),
\ee
i.e. $V^C_{k=0}=L(t_w)^d f(t/t_w)$, with $\lim_{t\to \infty}f(t/t_w)=$const.
This behavior has been derived in the sector of large $t-t_w$ but the scaling~(\ref{scalbelowx}) 
implies its general validity. A similar result, but for a somewhat different definition of $V^C$ is found in 
Ref.~\cite{c4sot}.
Going back to the data, the saturation for large $t$ predicted by the above arguments 
is better observed in $d=1$ (Fig.~\ref{d1neqbelow}) while in $d=2$, due to computer time
limitations, the 
data of Fig.~\ref{d2neqbelow} only show a tendency.

The data for $V^{C\chi}_{k=0,ag}(t,t_w)$ and ${\cal V}^\chi_{k=0,ag}(t,t_w)$
collapse with an exponent consistent with $a_{C\chi}=a_\chi=1/2$.
For large values of $(t-t_w)/t_w$, ${\cal V}^\chi_{k=0}$ grows as ${\cal V}^\chi\propto t^{1/2}$ 
while $V^{C\chi}$ approach a constant value, similarly to $V^C$.
In conclusion, our data show that $a_C=d/2$, $a_{C\chi}=a_\chi=1/2$,
and $\lambda _C=\lambda _{C\chi}=0$, $\lambda _{\chi}=1/2$ hold for $d=1,2$,
suggesting that this might be the generic behavior for all $d$~\cite{nota3d}.

\subsection{Time-re-parametrization invariance} \label{time}

In a series of papers \cite{arxiv/0109150} it was shown that the action describing the long time
slow dynamics of spin-glasses is invariant under a
re-parametrization of time $t\to h(t)$. 
Since $C$ and $\chi $ have the same scaling dimension the parametric
form $\chi (C)$ is also invariant under time reparametrizations.
Elaborating on this, it was claimed that
the long time physics of aging systems is characterized by Goldstone modes
in the form of slowly spatially varying reparametrizations $h_r(t)$,
similarly to spin waves in $O(N)$ models.
According to this physical interpretation, it was conjectured that 
fluctuating two-time functions measured locally 
by spatially averaging over a box of size $l$ centered on
${\bf r}$, i.e. $\widehat C _r(t,t_w)=\sum _i\widehat C _i(t,t_w)\theta (l-\vert i-r\vert)$
and similarly for $\widehat \chi _r(t,t_w)$, 
should fall
on the master-curve $\chi (C)$ of the average quantities in the
asymptotic limit where $t$ and $t_w$ are large. This was checked to be consistent
with numerical results for glassy models in Refs.~\cite{numtris}.
The choice of $l$ should be such that $l\sim R(t)$, where $R(t)$ is
the typical length over which the $h_r(t)$ variations occur.

Let us first observe that, with the definitions~(\ref{keq0}) and the discussion of 
$l$ at the beginning of Sec.~\ref{numbe}, the variances
$V_{k=0}^C$ and $V_{k=0}^\chi$ 
considered in this paper coincide with the variances of
the fluctuating quantities $\widehat C_r$, $\widehat \chi _r$
introduced above, provided that $l$ is the same in both cases.
Our results
for $V_{k=0}^X$, therefore, allow us to comment on this issue. 
Before doing so, however, let us recall once again that
${\cal V}^\chi$ is not the variance of the fluctuating
$\widehat \chi $.
Hence, from the analysis of ${\cal V}^\chi _{k=0}$ one cannot
directly infer the properties of $\widehat \chi $.
On the other hand, it is clear that
$\widehat \chi $ cannot fit {\it a priori} into the time-re-parametrization 
invariance scenario, since its variance contains the diverging terms
$K_i^\chi$ or $\tilde K_i^\chi$ of Eqs.~(\ref{bbbbb},\ref{kchi}),
according to Eqs.~(\ref{app1.6},\ref{divh}). Hence, the numerical results
contained in~\cite{numtris}, which are obtained by switching on the perturbation,
can only be consistent with that
scenario if a sufficiently large value of the perturbation $h$
is used in the simulations, so that the first contribution on the 
r.h.s. of Eq.~(\ref{kchi}) can be neglected.

The results of Sec.~\ref{quench} show that 
$\lim _{t_w\to \infty} \lim _{t\to \infty} {\cal V}^\chi_{k=0}/V^C_{k=0}=\infty $,
both in the quench to $T_c$ and below $T_c$. 
Since $V^{\chi}\ge {\cal V}^\chi$ (see Eq.~(\ref{app1.6}) 
or Eq.~(\ref{divh})) this implies that 
the fluctuations $\widehat \chi $ and $\widehat C$ cannot be constrained to
follow the $\chi (C)$ curve, at least in this particular order of 
the large time limits, as already noticed in~\cite{corbandolo}.
Hence the interpretation of~\cite{arxiv/0109150,numtris} 
cannot be strictly obeyed.
This may indicate either that the
symmetry $t\to h(t)$ is not obeyed in coarsening systems, as
claimed in~\cite{nontris}, or that its physical interpretation
misinterprets the effects of time-re-parametrization invariance
in phase-ordering kinetics.
Actually, the results of~\cite{corbandolo} show that at least
the limiting slope $X_\infty$ of $\chi (C)$ is encoded 
in the distribution of $\widehat C$ and $\widehat \chi $.
Whether this feature might be physically interpreted as a different
realization of time-re-parametrization invariance in coarsening system
is as yet unclear.

\section{Conclusions} \label{concl}

In this paper we have considered the fluctuations of two-time quantities
by studying their variances and the related second-order susceptibility. 
In doing that a first problem arises already at the level of their
definition. While $C_i$ is quite naturally
associated with the fluctuating quantity $[\sigma _i(t)-\langle\sigma _i(t)\rangle]
[\sigma_i(t_w)-\langle\sigma_i(t_w)\rangle]$,
for $\chi _i$ the situation is not as clear. Actually, referring
to the very meaning of a response function the straightforward 
way to associate a fluctuation to $\chi _i$ would be
the choice of Eq.~(\ref{pertfluc}) which is defined on a perturbed process. 
The quantity $\widehat \chi _i$ introduced
in this way, however, has 
diverging moments. A way out of this problem is to resort
to fluctuation-dissipation theorems. One may
enforce a relation between $\chi _i$ and the average of a fluctuating quantity
$\widehat \chi _i$, Eq.~(\ref{fdt2}), which holds also out
of equilibrium.
We have shown that the variances involving 
$\widehat \chi _i$ have a very weak dependence on the particular choice of this fluctuating part,
with the exception of the equal sites variance $V^\chi _{ii}$. 
For $i\neq j$, $V^{\chi}_{ij}$ is also a
second-order susceptibility ${\cal V}^\chi_{ij}$, which allows one to derive 
an equilibrium relation between variances, the SOFDT,
analogous to the FDT for the averages. Interestingly, 
the FDT and the SOFTD can be written in a rather similar form, namely 
Eqs.~(\ref{2.5}) and~(\ref{1quadratoa}),  expressing the vanishing 
of the first two moments of the quantity $\widehat D_i(t,t_w)$ defined in  
Eq.~(\ref{hatd}).
The SOFDT holds also for $i=j$ but in this case ${\cal V}^\chi_{ii}$
cannot be interpreted as a variance.

The SOFTD relates in
a quite natural way  ${\cal V}^\chi_{ij}$ to $V^C_{ij}$ promoting
the former to a role analogous to that advocated for the latter
in the context of disordered systems.
This suggests considering ${\cal V}^\chi_{ij}$ on an equal footing with
$V^C_{ij}$ and $V^{C\chi}_{ij}$ to study scaling behaviors and cooperativity.
This we have done in the second part of the Article, considering
ferromagnetic systems in and out of equilibrium. We have shown that
${\cal V}^\chi$, $V^C$ and $V^{C\chi}$
obey scaling forms involving the coherence length $\xi$ in equilibrium
or the growing length $L(t)$ after a quench, similarly to what is known
for $C$ and $\chi $. Our result are in good agreement with what is found 
analytically in the spherical model~\cite{annibale}.
They show that the time-re-parametrization invariance
scenario proposed for glassy dynamics does not hold strictly for ferromagnets,
as already guessed in~\cite{nontris,corbandolo}. This we find both 
in critical or in
sub-critical quenches, if the large-time limit is taken in the order
$\lim _{t_w\to \infty}\lim _{t\to\infty}$. Such a conclusion relies on the
fact that ${\cal V}_{k=0}^\chi /V_{k=0}^C\to \infty $ in this particular limit and,
hence, the fluctuations of $\widehat \chi $ cannot be exclusively triggered by those 
of $\widehat C$. Notice that this is true also in critical quenches where
$X_\infty $ is finite, showing that, quite obviously, a finite limiting effective
temperature does not guarantee that the scenario proposed in~\cite{arxiv/0109150,numtris}  
necessarily holds.  

\section*{Acknowledgments}

We thank Leticia Cugliandolo and A. Gambassi for discussions.

F.Corberi, M.Zannetti and A.Sarracino acknowledge financial support
from PRIN 2007 JHLPEZ ({\it Statistical Physics of Strongly Correlated
Systems in Equilibrium and Out of Equilibrium: Exact Results and 
Field Theory Methods}).

\section*{Appendix I} \label{app1}

In this Appendix we first discuss a possible definition of the fluctuating part of
$\chi _i$ in a perturbed process (namely, after Eq. (\ref{1.2})) and then show 
that for every choice of $\widehat \chi _i$ one obtains the same variances except
for $V^{\chi}_{ii}$.

\subsection{Definition of $\mathbf{\widehat \chi _i}$ in a perturbed process}

From Eq. (\ref{1.2}) one has
\begin{equation}
\langle \sigma _i\rangle_h=\langle \sigma _i\rangle+\sum _j\chi_{ij}(t,t_w)h_j(t_w),
\end{equation}
where now we consider a perturbing field switched on from $t_w$ onwards,
and $\chi _{ij}$ is the two-point susceptibility.
Using a random field with $\overline h_j=0$ and $\overline {h_ih_j}=h^2\delta _{ij}$ 
(where $\overline {\ldots}$ means an average over the field realizations)
one can single out the equal site susceptibility as \cite{barrat}
\begin{equation}
\chi _i(t,t_w)=\frac{1}{h^2}\overline{\langle \sigma _i(t)\rangle_h h_i(t_w)}.
\label{abcde}
\end{equation}
We stress here that, in doing so, for computing $\chi _i$ 
the perturbation does not need to be switched on only on the site $i$ as in Eq. (\ref{1.2}),
and this allows one to consider higher moments, such as the variances $V^{\chi }_{ij}$,
where the field must be switched on on both sites $i$ and $j$. 
Indeed one can introduce a (perturbed) fluctuating part of the susceptibility
as
\begin{equation}
\widehat \chi _i(t,t_w)=\frac{1}{h^2}\sigma _i(t)h_i(t_w),
\label{pertfluc}
\end{equation}
and the correlator
\begin{equation}
\langle\widehat{\chi_i}(t,t_w)\widehat{\chi_j}(t,t_w)\rangle=\frac{1}{h^4}\overline{\langle\sigma _i(t)
\sigma _j(t)\rangle_h h_i(t_w)h_j(t_w)}.
\label{corchih}
\end{equation}

\subsection{Independence of the variances of the choice of $\mathbf{\widehat \chi _i}$}

For the sake of simplicity, let us consider a discrete time dynamics with
two-time conditional probability given by

\begin{equation}
P(\sigma,t|\sigma',t_w)=\prod_{t'=t_w}^{t-1} w_h(\sigma(t'+1)|\sigma(t')),
\label{pazz}
\end{equation}
where $\sigma (t)$ is the configuration of the system at time $t$
and $w_h$ are the transition rates in the perturbed evolution.
The linear susceptibility can always be written in the form
\begin{equation}
\chi_i(t,t_w)=
\sum_{t'=t_w}^t\langle \sigma_i(t)a_i(t')\rangle 
\label{bert}
\end{equation}
with~\cite{Berthier}
\begin{equation}
a_i(t')=\left. \frac{\delta \ln w_h(\sigma(t'+1)|\sigma(t'))}{\delta h_i(t')}\right |_{h=0}
\end{equation}
from which the fluctuating susceptibility can be defined in terms of unperturbed quantities as
\begin{equation}
\widehat{\chi_i}(t,t_w)=\sigma_i(t)\sum_{t'=t_w}^t a_i(t').
\label{azz}
\end{equation}
Notice that $a_i$ depends on the particular form of the perturbed transition probabilities $w_h$.
Then, since for a given unperturbed model there is an arbitrarity in the choice of
the perturbed transition rates ~\cite{rlin,non_lin1&2}, one has
different definitions of $\widehat \chi _i$ and, in principle,
different $\chi _i$. However, as discussed 
in \cite{rlin,algoritmi}, once the average is taken in Eq. (\ref{bert}), all these
definitions are expected to yield essentially the same determination of $\chi _i$,
apart from very tiny differences which exactly vanish in equilibrium or in the large-time regime.
With the definition~(\ref{azz}) the following correlators can be built 
\begin{equation}
\langle\widehat{\chi_i}(t,t_w)\widehat{\chi_j}(t,t_w)\rangle=
\sum_{t'=t_w}^{t-1}\sum_{t''=t_w}^{t-1}\langle\sigma_i(t)\sigma_j(t)a_i(t')a_j(t'')\rangle,
\label{corchi}
\end{equation}
and
\begin{equation}
\langle\widehat{C}_i(t,t_w)\widehat{\chi_j}(t,t_w)\rangle=
\sum_{t'=t_w}^{t-1}\langle\sigma_i(t)\sigma_i(t_w)\sigma_j(t)a_j(t')\rangle.
\label{corcchi}
\end{equation}
Even though these quantities explicitly depend on the particular choice of $a_i$,
we show in the following that they can all be written as (non-linear) response functions,
which, therefore, are not expected to depend on the form of $a_i$, in the sense discussed above 
for $\chi _i$. 
Indeed, considering for simplicity a single spin dynamics, using Eqs.~(\ref{pazz}) and
proceeding analogously to the derivation of $\chi _i$ (Eq.~(\ref{bert})),
for $i\ne j$ one can compute the following response functions
\begin{equation}
R^{(2,2)}_{ij;ij}(t,t;t',t'')\equiv \left .\frac{\delta^2\langle \sigma_i(t)\sigma_j(t)\rangle_h}{\delta h_i(t')\delta h_j(t'')}
\right |_{h=0}
=\langle\sigma_i(t)\sigma_j(t)a_i(t')a_j(t'')\rangle
\label{rr22}
\end{equation}
and
\begin{equation}
R^{(3,1)}_{iij;j}(t,t_w,t;t')\equiv \left .\frac{\delta\langle \sigma_i(t)\sigma_i(t_w)\sigma_j(t)\rangle_h}{\delta h_j(t')}
\right |_{h=0}
=\langle\sigma_i(t)\sigma_i(t_w)\sigma_j(t)a_j(t')\rangle.
\label{rr31}
\end{equation}
Comparing Eqs.~(\ref{corchi},~\ref{corcchi}) with 
Eqs.~(\ref{rr22},~\ref{rr31}) one concludes 
that the correlators~(\ref{corchi},\ref{corcchi}) can be both related to
response functions the value of which, as for $\chi _i$,
are not expected to depend significantly on the choice of the form of the $w_h$
(and hence of $a_i$). The same holds, therefore, for the variances $V^\chi_{i,j}$ and $V^{C\chi}_{ij}$. Incidentally, Eqs. (\ref{corchi},\ref{rr22},\ref{chi2r2}) show also that 
$V^\chi_{ij}={\cal V}^\chi _{ij}$ for $i\ne j$. We stress that the above argument holds for every $ij$ for $V^{C\chi}_{ij}$
whilst it cannot be extended to the equal site variance $V^\chi _{ii}$, as we will show in Sec. \ref{app1c}.
Along the same lines, one can show that also the variances obtained with the perturbed fluctuating part~(\ref{pertfluc}) are related to the same response functions~(\ref{rr22},\ref{rr31}),
and hence take the same values. 
For instance, for the correlator~(\ref{corchih}), since 
\begin{eqnarray}
\lim_{h\to 0}\frac{1}{h^4}
\overline{\langle\sigma_i(t)
\sigma_j(t)\rangle _h h_i(t_w)h_j(t_w)}=
\sum _{t_1=t_w}^{t-1}\sum _{t_2=t_w}^{t-1}
\left . 
\frac{\delta^2 \langle \sigma _i(t)\sigma _j(t)\rangle_h}
{\delta h_{i}(t_1)\delta h_{j}(t_2)}\right |_{h=0} 
\label{nuova}
\end{eqnarray}
one has again
\begin{equation}
\langle\widehat \chi_i(t,t_w)\widehat{\chi_j}(t,t_w)\rangle=
\sum _{t_1=t_w}^{t-1}\sum _{t_2=t_w}^{t-1}
R^{(2,2)}_{ij;ij}(t,t;t_1,t_2).
\label{r222}
\end{equation}

\subsection{Equal sites} \label{app1c}

In order to discuss the behavior of $V^\chi _{ii}$ we compute explicitly
this quantity and ${\cal V}^\chi _{ii}$ making
the specific choice of $w_h$ which leads to Eq.~(\ref{1.5}). 
Using the second order fluctuation-dissipation relations derived in~\cite{non_lin1&2},
for Ising spins $R^{(2,2)}_{ij;ij}(t,t;t_1,t_2)$ can be rewritten as

\begin{eqnarray}
R^{(2,2)}_{ij;ij}(t,t;t_1,t_2)& = &
\frac{1}{4}  \Big \{ {\partial \over \partial t_1}{\partial \over \partial t_2}
\langle\sigma_i(t)\sigma_j(t)\sigma_i(t_1)\sigma_j(t_2)\rangle \nonumber \\
& - & {\partial \over \partial t_1}
\langle \sigma_i (t) \sigma_j (t) \sigma_i(t_1)
B_j(t_2)\rangle  \nonumber \\
& - & {\partial \over \partial t_2}
\langle \sigma_i (t) \sigma_j (t) B_i(t_1)
\sigma_j(t_2)\rangle  \nonumber \\
& + &  \langle \sigma_i (t) \sigma_j (t) B_i(t_1)
B_j(t_2)\rangle  \Big \} \nonumber \\
&+&\frac{1}{2}\delta(t_1-t_2)\delta_{ij}
\langle\sigma_i(t)^2 B_i(t_1)\sigma_i(t_1)\rangle.
\label{app1.3}
\end{eqnarray}
Using the property $\sigma _i^2=1$ the term 
$\langle\sigma_i(t)^2 B_i(t_1)\sigma_i(t_1)\rangle$
can be cast as 
$\frac{1}{2}\langle\sigma_i(t)^2 \tilde{B}_i(t_1)\rangle$, 
where $\tilde{B}_i=-\sum_{\sigma'}[\sigma'-\sigma]^2w(\sigma'|\sigma)$.
Writing this term in this form Eq. (\ref{app1.3}) holds generally 
for generic discrete or continuous variables. We will use this expression
in the following.
Integrating over $t_1$ and $t_2$ one obtains

\begin{eqnarray}
{\cal V}^\chi_{ij}&=&\frac{1}{4}\Big \{
\langle\sigma_i(t)\sigma_j(t)[\sigma_i(t)-\sigma_i(t_w)][\sigma_j(t)-\sigma_j(t_w)]\rangle
\nonumber \\
&-&\int_{t_w}^t dt_2 \langle\sigma_i(t)\sigma_j(t)[\sigma_i(t)-\sigma_i(t_w)]B_j(t_2)\rangle
\nonumber \\
&-&\int_{t_w}^t dt_1 \langle\sigma_i(t)\sigma_j(t)B_i(t_1)[\sigma_j(t)-\sigma_j(t_w)]\rangle
\nonumber \\
&+&\int_{t_w}^t dt_1\int_{t_w}^t dt_2 \langle\sigma_i(t)\sigma_j(t)B_i(t_1)B_j(t_2)\rangle
\Big\}\nonumber \\
&+&\frac{1}{4}\delta_{ij}\int_{t_w}^t dt_1 
\langle\sigma_i(t)^2 \tilde{B}_i(t_1)\rangle
-\chi_i(t,t_w)\chi_j(t,t_w).
\label{app1.4}
\end{eqnarray}

On the other hand, from the definitions~(\ref{1.5}) and~(\ref{1.71}) one has
\begin{eqnarray}
V^{\chi}_{ij}(t,t_w)&=&\langle\widehat{\chi}_i(t,t_w)\widehat{\chi}_j(t,t_w)\rangle
-\chi_i(t,t_w)\chi_j(t,t_w) \nonumber \\
&=&\frac{1}{4}\Big[\langle\sigma_i(t)\sigma_i(t)\sigma_j(t)\sigma_j(t)\rangle
-\langle\sigma_i(t)\sigma_i(t)\sigma_j(t)\sigma_j(t_w)\rangle \nonumber \\
&-&\int_{t_w}^t dt_1\langle\sigma_i(t)\sigma_i(t)\sigma_j(t)B_i(t_1)\rangle
-\langle\sigma_i(t)\sigma_j(t)\sigma_j(t)\sigma_i(t_w)\rangle
\nonumber \\
&+&\langle\sigma_i(t)\sigma_i(t_w)\sigma_j(t)\sigma_j(t_w)\rangle +
\int_{t_w}^t dt_1 \langle\sigma_i(t)\sigma_j(t)B_j(t_1)\sigma_i(t_w)\rangle \nonumber \\
&-&\int_{t_w}^t dt_1\langle\sigma_i(t)\sigma_j(t)\sigma_j(t)B_i(t_1)\rangle + 
\int_{t_w}^t dt_1 \langle\sigma_i(t)\sigma_j(t)B_i(t_1)\sigma_j(t_w)\rangle \nonumber \\
&+&\int_{t_w}^t dt_1 \int_{t_w}^t dt_2\langle\sigma_i(t)\sigma_j(t)B_i(t_1)B_j(t_2)\rangle
\Big]-\chi_i(t,t_w)\chi_j(t,t_w).
\label{app1.41}
\end{eqnarray}
This shows once again that in the case $i\ne j$ 
\begin{equation}
V^\chi_{ij}(t,t_w)={\cal V}^\chi_{ij}(t,t_w). 
\label{app1.5}
\end{equation}
For equal sites, on the other hand, one obtains from Eqs.~(\ref{app1.4}) and~(\ref{app1.41})
the following relation 
\be
V^{\chi}_{ij}(t,t_w)={\cal V}^\chi_{ij}(t,t_w)+K_i^{\chi}(t,t_w)\delta_{ij}, 
\label{app1.6}
\ee
where
\begin{equation}
{\cal V}^\chi_{ij}(t,t_w)= -\chi^2_i(t,t_w)-\Delta_i(t,t_w),
\end{equation}
(with $\Delta _i$ defined below Eq. (\ref{hyp}))
and
\be
K_i^{\chi}(t,t_w)=-\frac{1}{4}\int_{t_w}^t dt_1 
\langle\sigma_i(t)^2 \tilde{B}_i(t_1)\rangle.
\label{kappa1}
\ee
This quantity has been studied in specific models in \cite{algoritmi} and it is found 
to be positive and to diverge as $t-t_w$ increases. 
Finally, let us consider the equal site variance $V^\chi _{ii}$ in the case when the
perturbed definition~(\ref{pertfluc}) is used.
By forming products of $\widehat \chi  _{i}(t,t_w)$ one has
\be
V^\chi  _{ii}(t,t_w)=\lim _{h\to 0}\langle \left [\widehat {\delta \chi }_i (t,t_w)\right ]^2\rangle=
\tilde K_i^\chi,   
\label{divh}
\ee
with
\be
\tilde K_i^\chi(t,t_w)= T^2\lim _{h\to 0} h^{-2}-\chi _i^2(t,t_w)
\label{kchi}
\ee
This term 
diverges in the vanishing field limit. 
Since $V^{\chi}_{ii}$ is finite, this implies that 
$V^{\chi}_{ii}$ and $V^\chi_{ii}$ are necessarily different.

In conclusion, with every definition of the fluctuating part the variances
and ${\cal V}^\chi _{ij}$ turn out to be the same, with the exception of $V^{\chi}_{ii}$ which takes
different values.

\section*{Appendix II} \label{app2}

In this Appendix we derive the equilibrium relation~(\ref{1quadratoa})
among the variances, for $i\ne j$.
First, let us write explicitly the variances defined in Eqs.~(\ref{1.6}) and~(\ref{1.72})
with the help of Eq. (\ref{1.5})
\begin{eqnarray}
V^{C}_{ij}(t,t_w)&=&\langle\widehat C_i(t,t_w)\widehat C_j(t,t_w)\rangle
-C_i(t,t_w)C_j(t,t_w) 
\label{2.2}
\end{eqnarray}

\begin{eqnarray}
V^{C\chi}_{ij}(t,t_w)&=&\frac{1}{2}\Big[\langle\sigma_i(t)\sigma_i(t_w)\sigma_j(t)\sigma_j(t)\rangle-
\langle\sigma_i(t)\sigma_i(t_w)\sigma_j(t)\sigma_j(t_w)\rangle \nonumber \\
&-&\int_{t_w}^t dt_1 \langle\sigma_i(t)\sigma_j(t)B_j(t_1)\sigma_i(t_w)\rangle\Big]
-\langle\sigma_i(t)\rangle\langle\sigma_i(t_w)\rangle\chi_j(t,t_w) 
\nonumber \\
&-&C_i(t,t_w)\chi_j(t,t_w). 
\label{2.3}
\end{eqnarray}
The variance $V^{\chi}$ can be read from Eq.~(\ref{app1.41}).

In the following, we will use two properties involving the quantity $B_i$~\cite{glauber,rlin,non_lin1&2} 

\be
\langle B_i(t){\cal O}(t_1)\rangle
=\frac{\partial}{\partial t}\langle \sigma_i(t){\cal O}(t_1)\rangle
\qquad t>t_1,
\label{propB1}
\ee

\be
\langle [B_i(t)\sigma_j(t)+B_j(t)\sigma_i(t)]{\cal O}(t_1)\rangle=
\frac{\partial}{\partial t}\langle \sigma_i(t)\sigma_j(t){\cal O}(t_1)\rangle
\qquad t>t_1,
\label{propB2}
\ee
where ${\cal O}(t)$ is a generic observable.
In particular, at equilibrium, using time translation and time inversion invariance,
from Eq.~(\ref{propB1}) one has

\be
\langle {\cal O}(t)B_i(t_1)\rangle_{eq}=
-\frac{\partial}{\partial t_1}\langle {\cal O}(t)\sigma_i(t_1)\rangle_{eq}
\qquad t>t_1,
\label{Beq}
\ee
where we have introduced the notation $\langle\ldots\rangle_{eq}$ to indicate the equilibrium dynamics. 
This relation allows us to perform the integrals 
$\int_{t_w}^t dt_1\langle\sigma_i(t)\sigma_i(t)\sigma_j(t)B_i(t_1)\rangle$ and
$\int_{t_w}^t dt_1\langle\sigma_i(t)\sigma_j(t)\sigma_j(t)B_i(t_1)\rangle$
appearing in Eq.~(\ref{app1.41}). This yields

\begin{eqnarray}
V^{\chi}_{ij}(t,t_w)
&=&\frac{1}{4}\Big[3\langle\sigma_i(t)\sigma_i(t)\sigma_j(t)\sigma_j(t)\rangle_{eq}
-2\langle\sigma_i(t)\sigma_i(t)\sigma_j(t)\sigma_j(t_w)\rangle_{eq}
-2\langle\sigma_i(t)\sigma_j(t)\sigma_j(t)\sigma_i(t_w)\rangle_{eq}\nonumber \\
&+&\langle\sigma_i(t)\sigma_i(t_w)\sigma_j(t)\sigma_j(t_w)\rangle_{eq} 
+\int_{t_w}^t dt_1 \langle\sigma_i(t)\sigma_j(t)B_j(t_1)\sigma_i(t_w)\rangle_{eq} \nonumber \\
&+&\int_{t_w}^t dt_1\langle\sigma_i(t)\sigma_j(t)B_i(t_1)\sigma_j(t_w)\rangle_{eq} 
+\int_{t_w}^t dt_1 \int_{t_w}^t dt_2\langle\sigma_i(t)\sigma_j(t)B_i(t_1)B_j(t_2)\rangle_{eq}\Big]
\nonumber \\
&-&\chi_i(t,t_w)\chi_j(t,t_w).
\label{2.4}
\end{eqnarray}

Moreover, exploiting again the relation~(\ref{Beq}),
the double integral appearing into Eq.~(\ref{2.4}) can be rewritten as

\begin{eqnarray}
&&\int_{t_w}^t dt_1 \int_{t_w}^t dt_2\langle\sigma_i(t)\sigma_j(t)B_i(t_1)B_j(t_2)\rangle_{eq}
\nonumber \\
&=& \int_{t_w}^t dt_1 \int_{t_w}^{t_1} dt_2\left(-\frac{\partial}{\partial t_2}\right)
\langle\sigma_i(t)\sigma_j(t)B_i(t_1)\sigma_j(t_2)\rangle_{eq} \nonumber \\
&+&\int_{t_w}^t dt_2 \int_{t_w}^{t_2} dt_1\left(-\frac{\partial}{\partial t_1}\right)
\langle\sigma_i(t)\sigma_j(t)B_j(t_2)\sigma_i(t_1)\rangle_{eq} \nonumber \\
&=&-\int_{t_w}^t dt_1 \langle\sigma_i(t)\sigma_j(t)B_i(t_1)\sigma_j(t_1)\rangle_{eq}+
\int_{t_w}^t dt_1 \langle\sigma_i(t)\sigma_j(t)B_i(t_1)\sigma_j(t_w)\rangle_{eq} \nonumber \\
&-&\int_{t_w}^t dt_2 \langle\sigma_i(t)\sigma_j(t)B_j(t_2)\sigma_i(t_2)\rangle_{eq}+
\int_{t_w}^t dt_2 \langle\sigma_i(t)\sigma_j(t)B_j(t_2)\sigma_i(t_w)\rangle_{eq} \nonumber \\
&=&\int_{t_w}^t dt_1 \langle\sigma_i(t)\sigma_j(t)B_i(t_1)\sigma_j(t_w)\rangle_{eq}+
\int_{t_w}^t dt_1 \langle\sigma_i(t)\sigma_j(t)B_j(t_1)\sigma_i(t_w)\rangle_{eq} \nonumber \\
&+&\int_{t_w}^t dt_1 \frac{\partial}{\partial t_1}
\langle\sigma_i(t)\sigma_j(t)\sigma_j(t_1)\sigma_i(t_1)\rangle_{eq} \nonumber \\
&=&\int_{t_w}^t dt_1 \langle\sigma_i(t)\sigma_j(t)B_i(t_1)\sigma_j(t_w)\rangle_{eq} +
\int_{t_w}^t dt_1 \langle\sigma_i(t)\sigma_j(t)B_j(t_1)\sigma_i(t_w)\rangle_{eq} \nonumber \\
&+&\langle\sigma_i(t)\sigma_i(t)\sigma_j(t)\sigma_j(t)\rangle_{eq} 
-\langle\sigma_i(t)\sigma_j(t)\sigma_i(t_w)\sigma_j(t_w)\rangle_{eq},
\label{2.6}
\end{eqnarray}
where we have used the relation~(\ref{propB2}) to obtain the third equality.
Substituting this result into Eq.~(\ref{2.4}), one finally obtains

\begin{eqnarray}
V^{\chi}_{ij}(t,t_w)
&=&\frac{1}{4}\Big[4\langle\sigma_i(t)\sigma_i(t)\sigma_j(t)\sigma_j(t)\rangle_{eq}
-2\langle\sigma_i(t)\sigma_i(t)\sigma_j(t)\sigma_j(t_w)\rangle_{eq}
-2\langle\sigma_j(t)\sigma_j(t)\sigma_i(t)\sigma_i(t_w)\rangle_{eq} \nonumber \\ 
&+&2\int_{t_w}^t dt_1 \langle\sigma_i(t)\sigma_j(t)B_j(t_1)\sigma_i(t_w)\rangle_{eq}
+2\int_{t_w}^t dt_1\langle\sigma_i(t)\sigma_j(t)B_i(t_1)\sigma_j(t_w)\rangle_{eq} \Big] 
\nonumber \\
&-&\chi_i(t,t_w)\chi_j(t,t_w).
\label{2.5555}
\end{eqnarray}

Now, using the FDT $\chi_i(t,t_w)=C_i(t,t)-C_i(t,t_w)$ and assuming space translation invariance
$C_i(t,t_w)=C_j(t,t_w)$ and 
$\langle\sigma_i(t)\sigma_j(t)B_j(t_1)\sigma_i(t_w)\rangle_{eq}=
\langle\sigma_i(t)\sigma_j(t)B_i(t_1)\sigma_j(t_w)\rangle_{eq}$, from 
Eqs.~(\ref{2.2}),~(\ref{2.3}) and~(\ref{2.5555}) one obtains

\begin{equation}
V^C_{ij}(t,t_w)+2V^{C\chi}_{ij}(t,t_w)+V^{\chi}_{ij}(t,t_w)=
\langle\widehat C_i(t,t)\widehat C_j(t,t)\rangle_{eq}-C_i(t,t)C_j(t,t),
\end{equation}
which is the equilibrium relation~(\ref{1quadratoa}).

\section*{Appendix III}

In this Appendix we show that the quantity ${\cal V}^\chi_{ij}(t,t_w)$
at equal sites verifies the relation~(\ref{quadratoa}). 
In the case of Ising spins, since $R^{(2,2)}_{ij;ij}$ vanishes for $i=j$ by definition,
one immediately obtains ${\cal V}^\chi_{ii}=-\chi_i^2$, and, using the definitions~(\ref{2.2})
and~(\ref{2.3}) and the property $\sigma(t)^2\equiv 1$, one can easily check
that Eq.~(\ref{quadratoa}) holds.
In the case of continuous variables, in equilibrium, using the property~(\ref{Beq}),
from Eq.~(\ref{app1.4}) one has

\begin{eqnarray}
{\cal V}^\chi_{ii}(t,t_w)
&=&\frac{1}{4}\Big\{3\langle\sigma_i(t)^4\rangle_{eq}+
\langle\sigma_i(t)^2\sigma_i(t_w)^2\rangle_{eq}-4\langle\sigma_i(t)^3\sigma_i(t_w)\rangle_{eq} 
+4\int_{t_w}^t dt_1\langle\sigma_i(t)^2B_i(t_1)\sigma_i(t_w)\rangle_{eq} \nonumber \\
&-&2\int_{t_w}^t dt_1\langle\sigma_i(t)^2B_i(t_1)\sigma_i(t_1)\rangle_{eq} 
+\int_{t_w}^t dt_1\langle\sigma_i(t)^2 \tilde{B}_i(t_1)\rangle_{eq}\Big\}
-\chi_i(t,t_w)^2. 
\label{app3.1}
\end{eqnarray}
The quantities appearing in the last two terms in the braces at time $t_1$ 
can be rewritten as

\begin{equation}
-2B_i\sigma_i+\tilde{B}_i=-2\sum_{\sigma'}[\sigma'_i-\sigma_i]\sigma'_iw(\sigma'|\sigma)
+\sum_{\sigma'}[\sigma_i'^2+\sigma^2_i-2\sigma'_i\sigma_i]w(\sigma'|\sigma)=
\sum_{\sigma'}[-\sigma_i'^2+\sigma_i^2]w(\sigma'|\sigma)
\end{equation}
yielding 

\begin{equation}
\int_{t_w}^t dt_1\langle\sigma_i(t)^2[-2B_i(t_1)\sigma_i(t_1)+\tilde{B}_i(t_1)]\rangle_{eq} 
=\int_{t_w}^t dt_1 \frac{\partial}{\partial t_1}\langle\sigma_i(t)^2\sigma_i(t_1)^2\rangle_{eq}= 
\langle\sigma_i(t)^4\rangle_{eq}-\langle\sigma_i(t)^2\sigma_i(t_w)^2\rangle_{eq}. 
\end{equation}
Substituting this result into Eq.~(\ref{app3.1}), one finds

\begin{equation}
{\cal V}^\chi_{ii}(t,t_w)
=\langle\sigma_i(t)^4\rangle_{eq}-\langle\sigma_i(t)^3\sigma_i(t_w)\rangle_{eq} 
+\int_{t_w}^t dt_1\langle\sigma_i(t)^2B_i(t_1)\sigma_i(t_w)\rangle_{eq}-\chi_i(t,t_w)^2
\end{equation}
which coincides with Eq.~(\ref{quadratoa}), as can easily be checked recalling the
definitions~(\ref{2.2}) and~(\ref{2.3}).

\section*{Appendix IV} \label{app4}

In this Appendix we compute the large time limit of $V^C$, $V^{C,\chi}$ and ${\cal V}^\chi$,
starting from an equilibrium state ($t_w > t_{eq}$)
and taking the limit $t-t_w\to\infty$.
For the variance of the auto-correlation function one has 

\begin{eqnarray}
\lim_{t-t_w\to\infty}V^{C}_{ij}(t,t_w)&=&\lim_{t-t_w\to\infty}\Big[
\langle\widehat C_i(t,t_w)\widehat C_j(t,t_w)\rangle_{eq}
-C_i(t-t_w)C_j(t-t_w)\Big] \nonumber \\
&=&C_{ij,eq}(C_{ij,eq}+2m^2), 
\end{eqnarray}
where factorization at large $t-t_w$ has been used, and
$m=\langle\sigma_i\rangle_{eq}$ is the equilibrium 
magnetization.

For the covariance $V^{C\chi}_{ij}(t,t_w)$ one has

\begin{eqnarray}
\lim_{t-t_w\to\infty}V^{C\chi}_{ij}(t,t_w)&=&\lim_{t-t_w\to\infty}\Big\{
\frac{1}{2}\Big[C_i(t-t_w)-
\langle\sigma_i(t)\sigma_i(t_w)\sigma_j(t)\sigma_j(t_w)\rangle_{eq} \nonumber \\
&-&\int_{t_w}^t dt_1 \langle\sigma_i(t)\sigma_j(t)B_j(t_1)\sigma_i(t_w)\rangle_{eq}\Big] 
-m^2\chi_j(t,t_w)-C_i(t-t_w)\chi_j(t-t_w)\Big\} \nonumber \\
&=&\frac{1}{2}\Big[m^2-\langle\sigma_i\sigma_j\rangle^2_{eq}-2m^2(1-\langle\sigma_i\sigma_j\rangle^2_{eq})-
\lim_{t-t_w\to\infty}
\int_{t_w}^t dt_1 \langle\sigma_i(t)\sigma_j(t)B_j(t_1)\sigma_i(t_w)\rangle_{eq}\Big]
\nonumber \\
&-&m^2(1-m^2).
\label{Vcchieq}
\end{eqnarray}
The integral can be computed in the following way. Introduce an intermediate
time $t^*$ between $t$ and $t_w$ and take $t_w$, $t^*$ and
$t$ sufficiently far apart. Then one can write

\begin{eqnarray}
\int_{t_w}^t dt_1 \langle\sigma_i(t)\sigma_j(t)B_j(t_1)\sigma_i(t_w)\rangle_{eq} 
&=&\int_{t_w}^{t^*} dt_1 \langle\sigma_i(t)\sigma_j(t)\rangle_{eq}
\langle B_j(t_1)\sigma_i(t_w)\rangle_{eq} \nonumber \\
&+&
\int_{t^*}^t dt_1 \langle\sigma_i(t)\sigma_j(t)B_j(t_1)\rangle\langle\sigma_i(t_w)\rangle_{eq}. 
\label{intB}
\end{eqnarray} 
Using the property~(\ref{propB1}), the integral in the first term of the r.h.s 
can be rewritten as

\begin{eqnarray}
\int_{t_w}^{t^*} dt_1 \langle\sigma_i(t)\sigma_j(t)\rangle_{eq}
\langle B_j(t_1)\sigma_i(t_w)\rangle_{eq} &=&
\langle\sigma_i(t)\sigma_j(t)\rangle_{eq} \int_{t_w}^{t^*} dt_1 
\frac{\partial}{\partial t_1}\langle \sigma_j(t_1)\sigma_i(t_w)\rangle_{eq} \nonumber \\
&=& \langle\sigma_i(t)\sigma_j(t)\rangle_{eq}
[\langle\sigma_j(t^*)\sigma_i(t_w)\rangle_{eq}-\langle\sigma_j(t_w)\sigma_i(t_w)\rangle_{eq}]
\nonumber \\
&\rightarrow& m^2\langle\sigma_i\sigma_j\rangle_{eq}-\langle\sigma_i\sigma_j\rangle^2_{eq},
\label{intB1}
\end{eqnarray}
where in the last line the limit $t^*\to\infty$ has been taken.
Analogously, using the property~(\ref{Beq}), 
the second term on the r.h.s of Eq.~(\ref{intB}) can be computed, yielding

\be
-m(m-\langle\sigma_i\sigma_j\rangle_{eq}m).
\label{intB2}
\ee
Thus, replacing the integral appearing in Eq.~(\ref{Vcchieq}) with
Eqs.~(\ref{intB1}) and~(\ref{intB2}) one finds

\be
\lim_{t-t_w\to\infty}V^{C\chi}_{ij}(t,t_w)=-m^2C_{ij,eq}.
\ee

From Eq.~(\ref{2.5}), by means of analogous computations, one can easily check that

\be
\lim_{t-t_w\to\infty}{\cal V}^\chi_{ij}(t,t_w)=-C_{ij,eq}^2.
\ee


\begin{thebibliography}{99}
\bibitem {nota3d}
In order to check further
this point we have also computed the $V^C$, $V^{C,\chi}$ and ${\cal V}^\chi$ in $d=3$. 
Preliminary results seem to substantiate the 
dependence of the exponents discussed in the text.
\end{thebibliography}
\end{document}